\documentclass[twocolumn,english,prb,preprintnumbers,amsmath,amssymb,superscriptaddress]{revtex4}
\usepackage[latin9]{inputenc}
\usepackage{amsmath}
\usepackage{graphicx}

\makeatletter
\@ifundefined{textcolor}{}
{%
 \definecolor{BLACK}{gray}{0}
 \definecolor{WHITE}{gray}{1}
 \definecolor{RED}{rgb}{1,0,0}
 \definecolor{GREEN}{rgb}{0,1,0}
 \definecolor{BLUE}{rgb}{0,0,1}
 \definecolor{CYAN}{cmyk}{1,0,0,0}
 \definecolor{MAGENTA}{cmyk}{0,1,0,0}
 \definecolor{YELLOW}{cmyk}{0,0,1,0}
}

\usepackage{dcolumn}% Align table columns on decimal point
\usepackage{bm}% bold math
\usepackage{color}
\usepackage[final]{pdfpages}

\makeatother

\usepackage{babel}

\begin{document}

\preprint{preprint(\today)}

\title{Intriguing magnetism of the topological kagome magnet TbMn$_{6}$Sn$_{6}$}

%\pacs{74.20.Mn, 74.25.Ha, 74.70.Xa, 76.75.+i}

\author{C.~Mielke III}
%\thanks{These authors contributed equally to the paper.}
\affiliation{Laboratory for Muon Spin Spectroscopy, Paul Scherrer Institute, CH-5232 Villigen PSI, Switzerland}
\affiliation{Department of Physics, University of Z\"{u}rich, Winterthurerstrasse 190, Zurich, Switzerland}

\author{W. Ma}
%\thanks{These authors contributed equally to the paper.}
\affiliation{International Center for Quantum Materials,  School of Physics, Peking University, Beijing, China}

\author{V.~Pomjakushin}
\affiliation{Laboratory for Neutron Scattering  and Imaging, Paul Scherrer Institut, CH-5232 Villigen PSI, Switzerland}

\author{O.~Zaharko}
\affiliation{Laboratory for Neutron Scattering and Imaging, Paul Scherrer Institut, CH-5232 Villigen PSI, Switzerland}

\author{S.~Sturniolo}
\affiliation{Scientific Computing Department, Science \& Technology Facilities Council, Rutherford Appleton Laboratory, Didcot OX11 0QX, United Kingdom}

\author{X.~Liu}
\affiliation{Department of Physics, University of Z\"{u}rich, Winterthurerstrasse 190, Zurich, Switzerland}

\author{V.~Ukleev}
\affiliation{Laboratory for Neutron Scattering and Imaging, Paul Scherrer Institut, CH-5232 Villigen PSI, Switzerland}

\author{J.S.~White}
\affiliation{Laboratory for Neutron Scattering and Imaging, Paul Scherrer Institut, CH-5232 Villigen PSI, Switzerland}

\author{J.-X.~Yin}
\affiliation{Laboratory for Topological Quantum Matter and Spectroscopy, Department of Physics, Princeton University, Princeton, New Jersey 08544, USA}

\author{S.S.~Tsirkin}
\affiliation{Department of Physics, University of Z\"{u}rich, Winterthurerstrasse 190, Zurich, Switzerland}

\author{C.B.~Larsen}
\affiliation{Laboratory for Neutron Scattering and Imaging, Paul Scherrer Institut, CH-5232 Villigen PSI, Switzerland}

\author{T.A. Cochran}
\affiliation{Laboratory for Topological Quantum Matter and Spectroscopy, Department of Physics, Princeton University, Princeton, New Jersey 08544, USA}

\author{M.~Medarde}
\affiliation{Laboratory for Multiscale Materials Experiments, Paul Scherrer Institut, CH-5232 Villigen PSI, Switzerland}

\author{V.~Por\'{e}e}
\affiliation{Laboratory for Neutron Scattering and Imaging, Paul Scherrer Institut, CH-5232 Villigen PSI, Switzerland}

\author{D. Das}
\affiliation{Laboratory for Muon Spin Spectroscopy, Paul Scherrer Institute, CH-5232
Villigen PSI, Switzerland}

\author{R.~Gupta}
\affiliation{Laboratory for Muon Spin Spectroscopy, Paul Scherrer Institute, CH-5232
Villigen PSI, Switzerland}

\author{C.N. Wang}
\affiliation{Laboratory for Muon Spin Spectroscopy, Paul Scherrer Institute, CH-5232
Villigen PSI, Switzerland}

\author{J.~Chang}
\affiliation{Department of Physics, University of Z\"{u}rich, Winterthurerstrasse 190, Zurich, Switzerland}

\author{Z.Q.~Wang}
\affiliation{Department of Physics, Boston College, Chestnut Hill, MA, USA}

\author{R.~Khasanov}
\affiliation{Laboratory for Muon Spin Spectroscopy, Paul Scherrer Institute, CH-5232
Villigen PSI, Switzerland}

\author{T.~Neupert}
\affiliation{Department of Physics, University of Z\"{u}rich, Winterthurerstrasse 190, Zurich, Switzerland}

\author{A.~Amato}
\affiliation{Laboratory for Muon Spin Spectroscopy, Paul Scherrer Institute, CH-5232
Villigen PSI, Switzerland}

\author{L.~Liborio}
\affiliation{Scientific Computing Department, Science \& Technology Facilities Council, Rutherford Appleton Laboratory, Didcot OX11 0QX, United Kingdom}

\author{S.~Jia}
\affiliation{International Center for Quantum Materials,  School of Physics, Peking University, Beijing, China}
\affiliation{CAS Center for Excellence in Topological Quantum Computation, University of Chinese Academy of Science, Beijing, China}

\author{M.Z.~Hasan}
%\email{mzhasan@princeton.edu}
\affiliation{Laboratory for Topological Quantum Matter and Spectroscopy, Department of Physics, Princeton University, Princeton, New Jersey 08544, USA}

\author{H.~Luetkens}
%\email{hubertus.luetkens@psi.ch}
\affiliation{Laboratory for Muon Spin Spectroscopy, Paul Scherrer Institute, CH-5232 Villigen PSI, Switzerland}

\author{Z.~Guguchia}
\email{zurab.guguchia@psi.ch}
\affiliation{Laboratory for Muon Spin Spectroscopy, Paul Scherrer Institute, CH-5232 Villigen PSI, Switzerland}

%\newpage
%\end{abstract}

\begin{abstract}

\textbf{Magnetic topological phases of quantum matter are an emerging frontier in physics and material science \cite{JXYin1,GuguchiaNat,JXYin2,LYe,Mazin,THan,SYan,Keimer,WangZhang,HasanKane}. Along these lines, several kagome magnets have appeared as the most promising platforms. Here, we explore magnetic correlations in the transition-metal-based kagome magnet TbMn$_{6}$Sn$_{6}$ using muon spin rotation, combined with local field analysis and neutron diffraction \cite{TbNature}. Our results show that the system exhibits an out-of-plane ferrimagnetic structure $P6/mm'm'$ (comprised by Tb and Mn moments) with slow magnetic fluctuations below $T_{\rm C2}$~=~320~K. These fluctuations exhibit a slowing down below $T_{\rm C1}^{*}$~${\simeq}$~120~K, and we see the formation of static patches with ideal out-of-plane order below $T_{\rm C1}$~${\simeq}$~20~K which grow in a volume with decreasing temperature.  The appearance of the static patches has a similar onset to the interesting phenomenon such as spin-polarized Dirac dispersion with a large Chern gap and topological edge states. We further show that the temperature evolution of the anomalous Hall conductivity (AHC) is strongly influenced by the low temperature magnetic crossover. Our presented experimental results show that the onset of the topological electronic properties tied to the Dirac band is promoted only by true static out-of-plane ferrimagnetic order in TbMn$_{6}$Sn$_{6}$ and is washed out by the slow magnetic fluctuations above $T_{\rm C1}$~${\simeq}$~20~K. Remarkably, hydrostatic pressure of 2.1 GPa stabilises static out-of-plane topological ferrimagnetic ground state in the whole volume of the sample. Therefore the exciting perspective arises of a magnetic system in which the topological response can be controlled, and thus explored, over a wide range of parameters.}

\end{abstract}
\maketitle

\section{INTRODUCTION}

 With distinguished symmetry and associated geometrical frustration, the kagome lattice can host peculiar states including flat bands \cite{JXYin1}, Dirac fermions \cite{LYe,JXYin2}  and spin liquid phases \cite{THan,SYan}. Magnetic kagome materials are an ideal setting in which strongly correlated topological electronic states may emerge \cite{Keimer,Mazin,WangZhang,HasanKane,Wen,TbNature,GuguchiaNat}. In particular, transition-metal-based kagome magnets \cite{FelserCSS,GuguchiaNat,THan,JXYin1,LYe,Fenner,SYan,JXYin2,Nakatsuji,Shuang,Nayak,Ghimire} are emerging as outstanding candidates for such states, as they feature both large Berry curvature fields and unusual magnetic tunability. As an example, we established Co$_{3}$Sn$_{2}$S$_{2}$ as a material that hosts frustrated magnetism in the kagome lattice \cite{GuguchiaNat} and find that the volume-wise magnetic competition drives the thermodynamic or quantum evolution of the intrinsic AHC, thereby tuning its topological state. A number of nontrivial magnetic phases and a large topological Hall effect was also observed in another  rare earth-transition metal based system YMn$_{6}$Sn$_{6}$ \cite{Ghimire,Pengcheng,Dally}. A new nematic chirality mechanism, which originates in frustrated interplanar exchange interactions that trigger strong magnetic fluctuations, was discussed as the reason for the topological Hall effect.  This suggests that detailed experimental studies to understand the magnetic fluctuations as well as the static magnetic structure of these systems is necessary to reveal the coupling between relativistic electrons and the magnetic properties.

 In the so-called 166 materials, TbMn$_{6}$Sn$_{6}$ contains the heavy rare earth Tb and crystallizes in a HfFe$_{6}$Ge$_{6}$-type structure (space group $P6$/$mmm$) composed of hexagonal Tb layers containing Sn atoms and Mn kagome nets, stacked in the sequence -Mn-Tb-Mn-Mn-Tb-Mn- along the $c$-axis (Figure~2a) \cite{Venturini,Goto,Zhang2005,Malaman,Clatterbuck,Vulliet,Zhang}. Structurally, compared with the Co$_{3}$Sn$_{2}$S$_{2}$ system consisting of a Co$_{3}$Sn kagome layer and a Sn$_{2}$ honeycomb layer, the $R$Mn$_{6}$Sn$_{6}$ system features a pristine Mn$_{3}$ kagome layer (Fig. 2a), as the Sn atoms are pushed away from the kagome layer by the chemical pressure from the $R$ atoms. TbMn$_{6}$Sn$_{6}$ has different Tb and Mn magnetic sublattices. Previous neutron diffraction studies indicated that this compound exhibits in-plane ferrimagnetic ordering between Tb and Mn sublattices below the Curie temperature $T_{\rm C}$~=~423~K \cite{Venturini, Malaman}, due to the strong Tb-Mn exchange interaction. It was also reported that at $T_{\rm C2}$~=~320~K \cite{Venturini, Malaman}, a spin reorientation transition occurs and the easy magnetization direction changes from the $ab$-plane at high temperatures to the $c$-axis at low temperatures. From the electronic point of view, TbMn$_{6}$Sn$_{6}$ exhibits Shubnikov-de Haas quantum oscillations with nontrivial Berry phases at relatively low fields (from ${\sim}$~7 T), a large AHC (0.14~$e^{2}$/h per Mn kagome layer) arising from Berry curvature fields, and quasi-linear (${\propto}$~$H^{1.1}$) magnetoresistance (MR) likely resulting from linearly dispersive electrons \cite{TbNature,Shuang}. Moreover, TbMn$_{6}$Sn$_{6}$ was found to demonstrate a bulk-boundary correspondence between the Chern gap and the topological edge state, as well as Berry curvature field correspondence of Chern-gapped Dirac fermions \cite{TbNature,Shuang}. Thus, it is identified as a promising topological magnetic system \cite{TbNature,Shuang}. Other than the identified high static magnetic ordering temperatures of TbMn$_{6}$Sn$_{6}$, microscopic studies of spin fluctuations and its tunability remain largely unexplored.

  Here we utilize high resolution muon-spin rotation (${\mu}$SR), a very powerful local magnetic probe, in combination with magnetization, powder and single crystal neutron diffraction to systematically characterize the phase diagram, uncovering intriguing slow magnetic fluctuations in TbMn$_{6}$Sn$_{6}$ in a wide temperature range down to 1.7~K. These fluctuations slow down below $T_{\rm C1}^{*}$~${\simeq}$~120~K and form static patches with an ideal out-of-plane ferrimagnetic order below $T_{\rm C1}$~${\simeq}$~20~K. 
  %which \textcolor{red}{differ from the disordered high-temperature magnetic state deduced from local field calculations.} 
  %which is suggested to be different and more disordered than the ideal out-of-plane ferrimagnetic state, observed at higher temperatures. 
  $T_{\rm C1}$ is considered as a magnetic crossover temperature signaling a slowing down of magnetic fluctuations rather than being regarded as a true phase transition temperature. We also find that the AHC does not follow the temperature evolution of the static Tb and Mn moments, but it is strongly influenced by the observed low-temperature magnetic crossover. Moreover, the large linear MR and the quantum oscillations, which are related to the quantized Landau fan structure featuring a spin-polarized Dirac dispersion with a large Chern gap, appear below $T_{\rm C1}^{*}$ and $T_{\rm C1}$, respectively. Thus, an intimate coupling between the novel low-$T$ magnetic crossover and the quantum-limit Chern gapped phase of TbMn$_{6}$Sn$_{6}$ is found.

%%%%%%%%%%%%%%%%%%%%%%%%%%%%%%%%%%%%%%%%%%%%%%%%%%%%%%%%%%
\begin{figure*}[t!]
\includegraphics[width=1.0\linewidth]{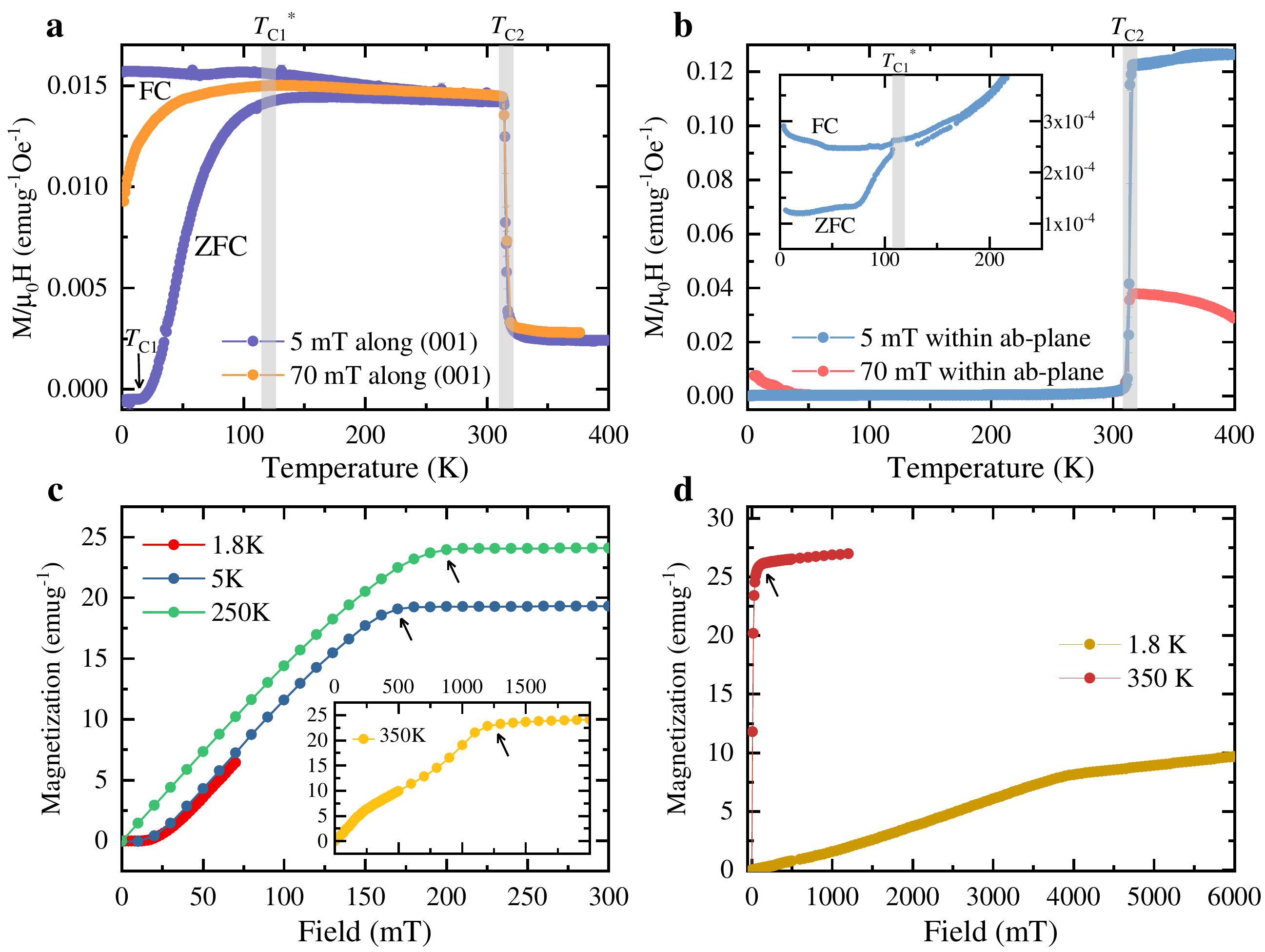}
\vspace{-0.5cm}
\caption{ \textbf{Temperature and field dependent macroscopic magnetic properties.}
(a-b) The temperature dependence of the zero-field cooled (sample was cooled down to the base-$T$ in zero magnetic field and the measurements were done upon warming) and field-cooled (the sample was cooled down to the base-$T$ in an applied magnetic field and the measurements were done upon warming) magnetization, measured in an field of 5 mT and 70 mT, applied along the (001) direction (a) and applied within the $ab$-plane (b). Arrows mark the magnetic transition temperatures $T_{\rm C1}$, $T_{\rm C1}^{*}$ and $T_{\rm C2}$. (c-d) The field dependence of the zero-field cooled magnetization, recorded at various temperatures with the field applied along the (001) direction (c) and within the $ab$-plane (d). }
\label{fig1}
\end{figure*}
%%%%%%%%%%%%%%%%%%%%%%%%%%%%%%%%%%%%%%%%%%%%%%%%%%%%%%%%%%

\section{RESULTS}

\subsection{Macroscopic magnetic properties}

 Temperature- and field-dependent magnetization experiments were performed in zero-field-cooled (ZFC) and field-cooled (FC) conditions for low applied fields of ${\mu}_{0}H$~=~5~mT and 70~mT, as shown in Figs.~1a~and~b. The field was applied both in-plane (Fig.~1b) and out-of-plane, along the crystallographic $c$-axis (Fig.~1a). We observe a large and sharp transition occurring at $T_{\rm C2}$~${\simeq}$~320~K, corresponding to the spin reorientation transition noted in previous neutron diffraction studies \cite{Venturini, Malaman}. This marks the transition between a high-temperature in-plane magnetic phase to a low-temperature ferrimagnetic phase with magnetic moments oriented out-of-plane, along the crystallographic $c$-axis. More importantly, when the magnetic field was applied along the $c$-axis, we observed a large reduction of zero-field cooled susceptibility ${\chi}_{\rm ZFC}$ below $T_{\rm C1}^{*}$ ${\simeq}$ 120~K and settling into a negligibly small slightly diamagnetic plateau below $T_{\rm C1}$ ${\simeq}$ 20~K. On the other hand, the field cooled susceptibility ${\chi}_{\rm FC}$ shows a weak temperature dependence across 120~K and down to the base-$T$. This gives rise to a much larger hysteresis in TbMn$_{6}$Sn$_{6}$ below 120~K than in the $T$-range of 120~K~--~320~K. The plateau completely disappears with the application of higher magnetic fields, as it is not visible with the application of even a modest field of 70~mT. The reduction of ${\chi}_{\rm ZFC}$ is also less pronounced in 70~mT.  The susceptibility, measured for a field applied in the $ab$-plane, is largely insensitive to the low-T transition (see Fig. 1b). So, the large difference between the FC and ZFC response is not caused by the appearance of the in-plane structure but, rather, is consistent with the scenario that different out-of-plane ferrimagnetic domains tend to cancel out (anti-align) after ZFC. If we FC the sample even in low fields, then the domains align. This suggests that magnetic states above and below 120~K are $c$-axis aligned, but slightly different from each other. This conclusion is also substantiated by the field dependent measurements of ZFC magnetization at $T$ = 1.8~K, 250~K, and at 350~K, measured for the field applied along the $c$-axis (Figure 1c) and along the $ab$-plane (Figure~1d). The critical fields above which the magnetization plateau appears are similar for $T$ = 1.8~K and 250~K. The only difference is that the data collected at 1.8~K show a plateau up to 12~mT, before increasing and then entering the linear regime, implying that the plateau, which is a characteristic feature for the low-$T$ state, is easily destroyed by weak applied fields. For H ${\parallel}$ $c$, the saturated magnetization does not occur at 350~K until a relatively high applied magnetic field of ${\simeq}$ 1.2~T (inset of Figure~1c), much higher than 175-200~mT at 5~K and 250~K (Fig. 1c). Conversely, when the field is applied along the $ab$-plane, saturation is almost immediately achieved upon application of 40~mT at 350~K, while at the base temperature the plateau was never achieved even under maximum applied field (7~T). These results are compatible with the in-plane ferrimagnetic order at 350~K.%same

%%%%%%%%%%%%%%%%%%%%%%%%%%%%%%%%%%%%%%%%%%%%%%%%%%%%%%%%%%
\begin{figure*}[t!]
\includegraphics[width=1.0\linewidth]{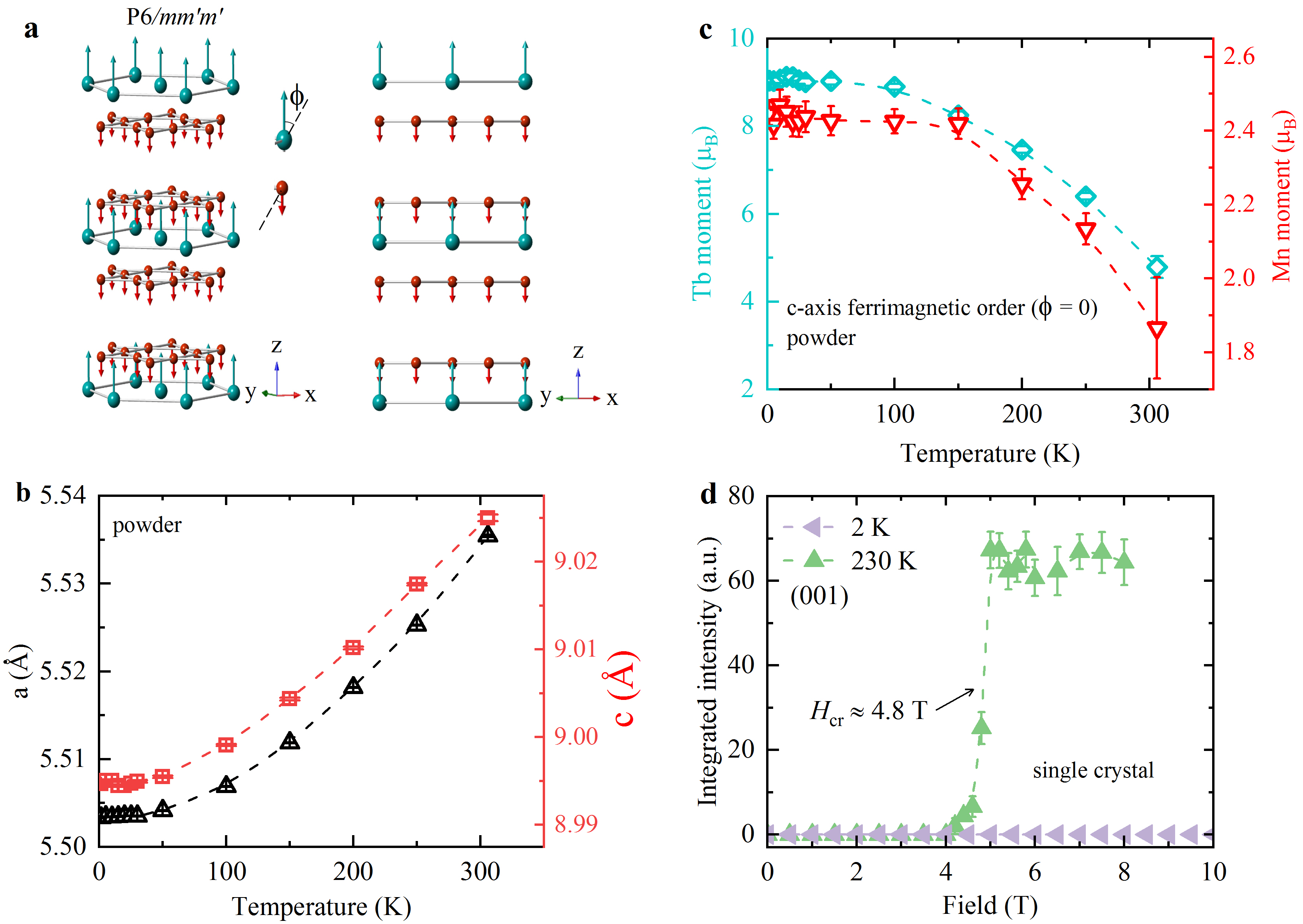}
\vspace{-0.2cm}
\caption{ \textbf{Crystal and magnetic structures of TbMn$_{6}$Sn$_{6}$.}
(a) Magnetic structure of TbMn$_{6}$Sn$_{6}$. ${\Phi}$ is the deviation angle from the $c$-axis. ${\Phi}$ = $0^{\circ}$ indicates out-of-plane ferrimagnetic order and ${\Phi}$ = $90^{\circ}$ indicates the in-plane ferrimagnetic order. The Mn atoms construct a kagome lattice (red middle size circles), while the Tb atoms (turquoise larger circles) form a honeycomb structure.
(b) The temperature dependence of the lattice constants $a$ and $c$ in TbMn$_{6}$Sn$_{6}$. (c) The temperature dependence of the terbium and manganese magnetic moments. The error bars represent the s.d. of the fit parameters.
(d) The field dependence of the (001) peak, recorded at 2 K and 130 K.} 
%(e-g) Comparison between calculated and observed intensities of the diffraction peaks at three different temperatures: 2 K, 70 K, and 250 K. The black line denotes perfect agreement between observed and predicted intensities at 250 K, which is not the case at 70 K and 2 K.}
\label{fig2}
\end{figure*}
%%%%%%%%%%%%%%%%%%%%%%%%%%%%%%%%%%%%%%%%%%%%%%%%%%%%%%%%%%

\subsection{Determination of magnetic structure}

 The temperature dependence of the sample magnetization clearly uncovers a novel transition/crossover at $T_{\rm C1}^{*}$~${\simeq}$~120~K. In order to characterize this novel transition, neutron scattering experiments were performed from 2~-~250~K using high resolution neutron powder \cite{Schefer,Fischer} and single crystal diffraction (see the Supplementary Information). From powder diffraction, the crystal structure was well refined by the Rietveld method, employing a hexagonal lattice structure in the space group $P6/mmm$, No.~191. Lattice constants $a$ and $c$ show a smooth monotonous decrease with decreasing temperature and a plateau below 50~K, as depicted in Fig. 2b. To solve the magnetic structure, the collected diffraction patterns were first refined with powder matching via the LeBail method with FullProf \cite{Rodriguez}, which confirmed the propagation vector at $k$~=~0 found by previous studies \cite{Venturini, Malaman}. Symmetry analysis shows that the maximal most symmetric subgroup mGM2+ ($P6/mm'm'$ No.~191.240) fits the data from 2~K to 250~K best, with a similar $\chi$$^{2}$ to the fit obtained via the LeBail method, implying that the model can hardly be improved. The temperature dependence of the estimated magnetic moments of Tb and Mn are shown in Fig.~2b. They both increase monotonically with decreasing temperature down to 100~K and stay nearly constant below this temperature.

 To further explore the low temperature magnetic structure, single crystal neutron diffraction measurements were performed (see Figure~2) both in the ordered state and in the paramagnetic regime ($T~\simeq~435~K$). The ideal out-of-plane ferrimagnetic $c$-axis aligned $P6/mm'm'$ structure has the best fit for the data collected at 150 K and 300 K;  however, at the lowest temperature, 5~K, the out-of-plane ferrimagnetic $P2/m$, No.~10.42 achieved a slightly better fit to the data (see Supplemental Information). This lower-symmetry structure allows for three different Mn sites and a mixing of two irreducible representations. Both $P6/mm'm'$ and $P2/m$ are characterised by a perfectly $c$-axis aligned structure (see Supplemental Information). Any incommensurability or out-of-plane canting of the magnetic moments were excluded by additional state-of-the-art Small Angle Neutron Scattering (SANS) measurements over a broad range of momentum transfer.

 At 2~K, the magnetic order stays $c$-axis aligned under the in-plane magnetic field as high as 10~T (see Fig.~2d). At 230~K, the spin reorientation transition from the $c$-axis to the $ab$-plane is induced, signaled by the abrupt emergence of the (001) peak at a critical in-plane magnetic field of 4.9~T (see Fig.~2d).

%%%%%%%%%%%%%%%%%%%%%%%%%%%%%%%%%%%%%%%%%%%%%%%%%%%%%%%%%%
\begin{figure*}[t!]
\includegraphics[width=1.0\linewidth]{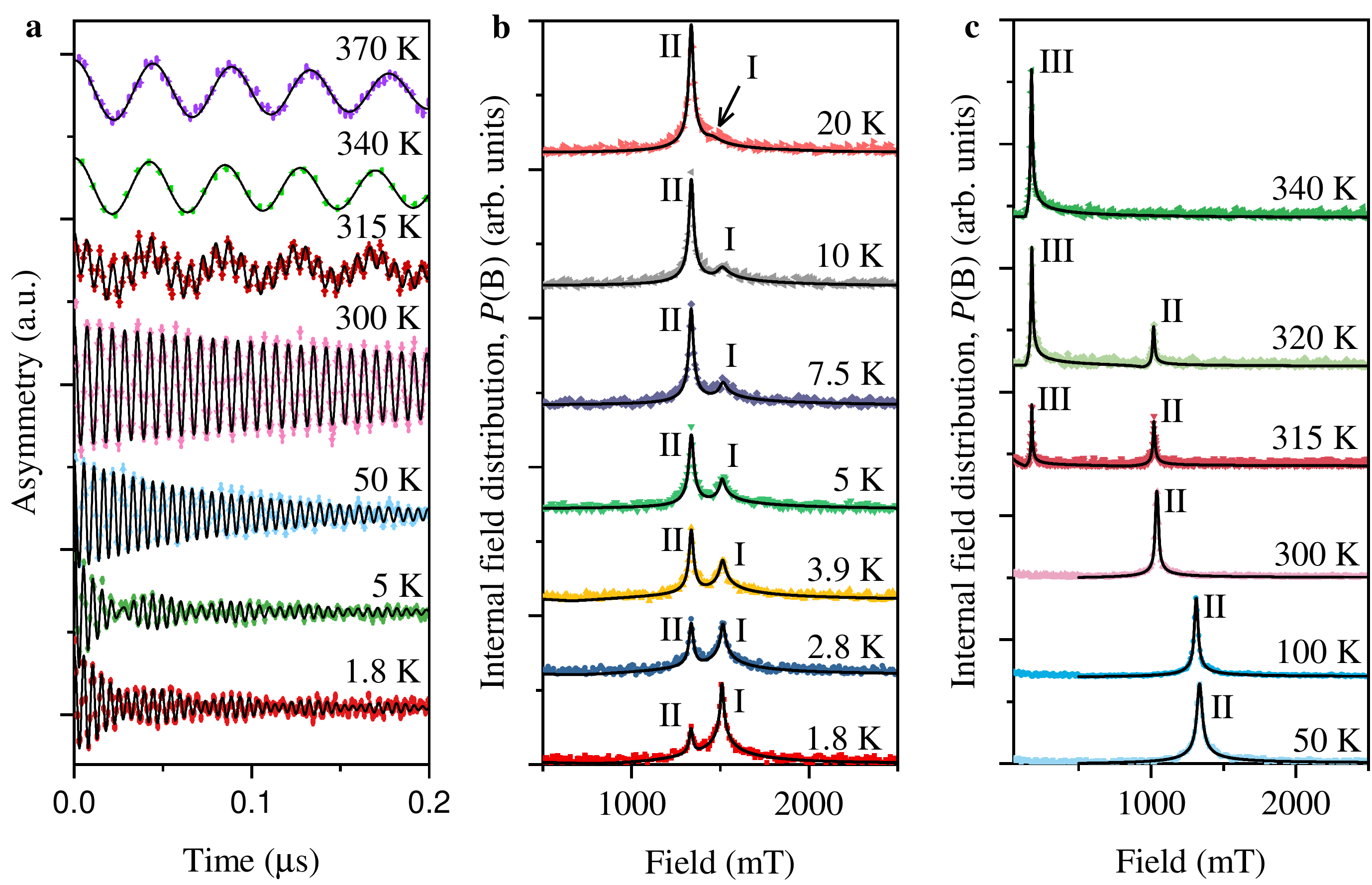}
\vspace{-0.5cm}
\caption{ \textbf{Zero-field ${\mu}$SR signals for TbMn$_{6}$Sn$_{6}$.}
Zero field time spectra, recorded at various temperatures in the temperature range between 1.8 K to 370 K. The solid lines are the fit of the data using Eq. 2. s. Error bars are the s.e.m. in about 10$^{6}$ events. The error of each bin count n is given by the s.d. of n. The errors of each bin in $A$($t$) are then calculated by s.e. propagation. (b-c) Fourier transform amplitudes of the oscillating components of the ${\mu}$SR time spectra, indicating the internal field distribution $P(B)$, as a function of temperature.}
\label{fig3}
\end{figure*}
%%%%%%%%%%%%%%%%%%%%%%%%%%%%%%%%%%%%%%%%%%%%%%%%%%%%%%%%%%%%

%%%%%%%%%%%%%%%%%%%%%%%%%%%%%%%%%%%%%%%%%%%%%%%%%%%%%%%%%%%%
\begin{figure*}[t!]
\includegraphics[width=1.0\linewidth]{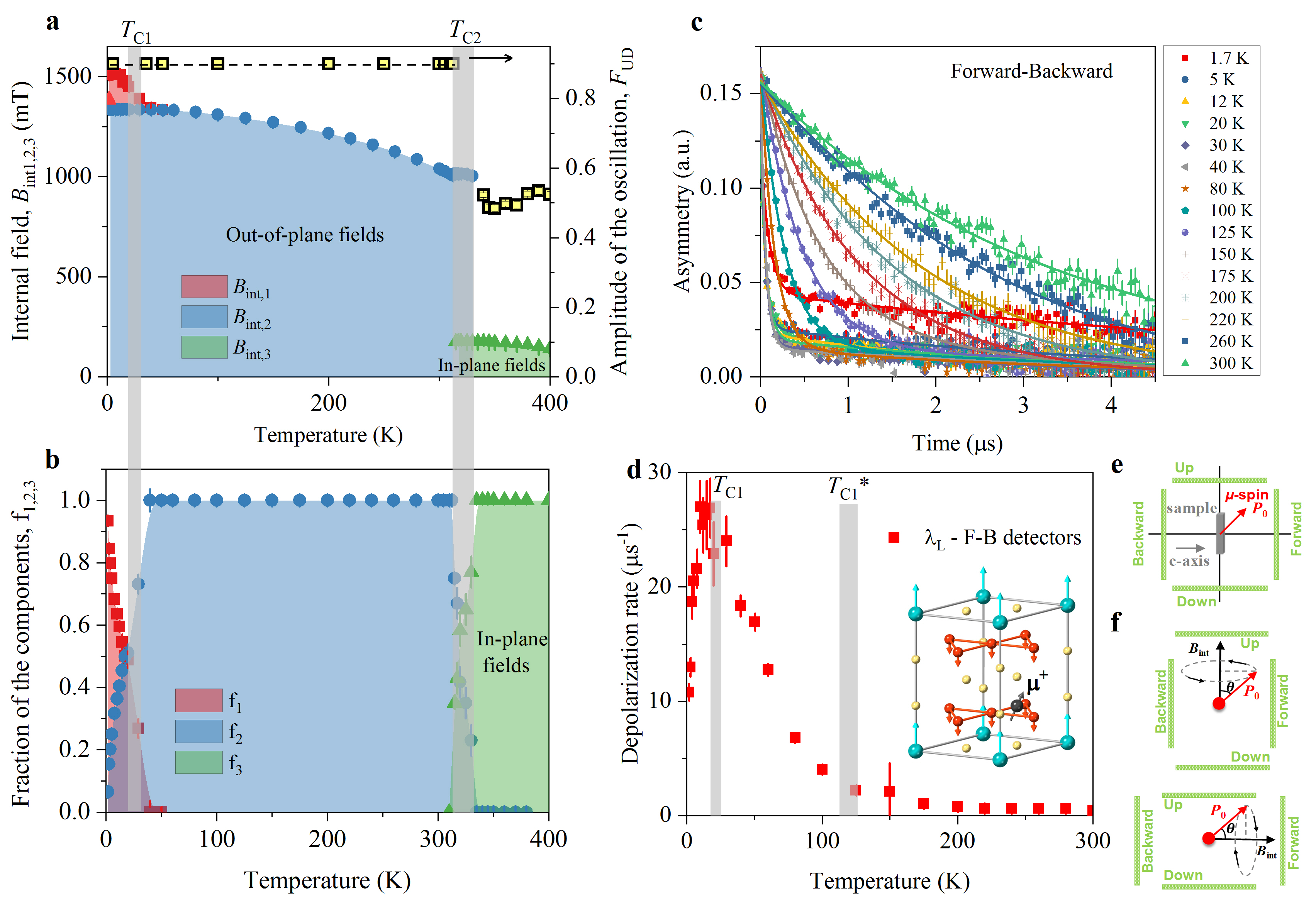}
\vspace{0cm}
\caption{ \textbf{Static order and dynamic fluctuations in TbMn$_{6}$Sn$_{6}$.}
(a) The temperature dependences of the internal magnetic fields for the three magnetic components. Vertical lines mark the critical temperatures $T_{\rm C1}$ and $T_{\rm C2}$.  $T_{\rm C2}$ is the transition temperature from high temperature low field to low temperature high field component, while $T_{\rm C1}$ is the transition temperature, below which two high field components, exhibiting volume wise competition, are observed. The error bars represent the s.d. of the fit parameters. Right axis depicts the amplitude of the oscillating component of the ${\mu}$SR signal from up-down (U-D) positron detectors. (b) The temperature dependences of the relative volume fractions ($f_{\rm 1}$, $f_{\rm 2}$, $f_{\rm 3}$) of the three magnetically ordered regions. (c) Zero-field ${\mu}$SR signals from Forward-Backward (F-B) positron detectors, recorded at various temperatures. 
(d) The temperature dependence of dynamic depolarization rate of the ${\mu}$SR signal, measured in F-B positron detectors. Arrows mark the magnetic transition temperature $T_{\rm C1}$ and the temperature $T_{\rm C1}^{*}$ for the onset of visible magnetic fluctuations. Inset of panel (d) shows the muon stopping site within the structure of TbMn$_{6}$Sn$_{6}$. (e) A schematic overview of the experimental setup for the muon spin forming $45^{\circ}$ with respect to the $c$-axis of the crystal. The sample was surrounded by four detectors: Forward (F), Backward (B), Up (U) and Down (D). (f) Schematic illustration of the muon spin precession around the internal magnetic field for two cases: (top) The field is perpendicular to the $c$-axis and points towards the U-detector. ${\theta}$ is the angle between the magnetic field and the muon spin polarization at $t$ = 0. (bottom) The field is parallel to the $c$-axis of the crystal and points towards the F-detector.}
\label{fig4}
\end{figure*}
%%%%%%%%%%%%%%%%%%%%%%%%%%%%%%%%%%%%%%%%%%%%%%%%%%%%%%%%%%

\subsection{Microscopic details of static and dynamic magnetic state}

  To gain further insight into the intriguing magnetic properties of TbMn$_{6}$Sn$_{6}$, we employed the ${\mu}$SR technique, which serves as an extremely sensitive local probe for detecting microscopic details of the static magnetic order, ordered magnetic volume fraction, and magnetic fluctuations. The local probe feature makes ${\mu}$SR a perfect complementary technique to neutron diffraction and magnetization measurements.

Fig.~3a displays representative zero-field (ZF) ${\mu}$SR time spectra for TbMn$_{6}$Sn$_{6}$ taken at various temperatures in the range from 1.8~K~--~400~K. ZF ${\mu}$SR spectra reveal coherent oscillations of the muon spin, indicating the existence of a well-defined internal field at the muon stopping sites in the sample in the whole investigated temperature range. This signal is expected for well defined  long-range magnetic order. Interestingly, either a single or a superposition of two distinct precession frequencies can be clearly seen in the ${\mu}$SR spectra. To better visualize the magnetic response, we show the Fourier transform amplitudes of the oscillating components of the ${\mu}$SR time spectra as a function of temperature (Fig.~3b and c), which is a measure of the probability distribution of internal fields sensed by the muon ensemble.
We observe only one low field component below 400~K (component III in Fig.~3c), then a coexistence of a low and high field components (component III and component II, respectively) between 330~K and 315~K (Fig.~3c), and the high field component persists down to the base-$T$ (Fig.~3b-c). Remarkably, below 20 ~K a second higher component appears (component I), which becomes the dominant component at the base temperature of 1.7~K (Figure 2b). The spectra below 20~K are also characterized by loss of a few percent of the initial asymmetry.

The temperature dependences of the internal magnetic fields and the relative fractions of the three internal field components for the single crystal of TbMn$_{6}$Sn$_{6}$ are shown in Figures 4a and b, respectively. There is a sharp and large (by a factor of 6) increase of the internal field across the high-temperature in-plane ferrimagnetic to a low-temperature $c$-axis ferrimagnetic phase transition temperature $T_{\rm C2}$ ${\simeq}$ 320~K, as shown in Figure~4a. However, there is a temperature range of about 15~K (marked by a vertical grey line) in which both in-plane and $c$-axis ferrimagnetic phases coexist in the sample but are macroscopically phase separated (see Figure~4b), pointing towards a first-order nature of the phase transition at $T_{\rm C2}$. Upon lowering the temperature below 310~K, the internal field for the out-of-plane component monotonously decreases down to the lowest temperature of 1.7~K. The internal field of the additional component monotonously increases below 25~K and saturates below 10~K. As can be clearly seen in Figure~4b, the fraction of component I increases at the cost of component II. Component I eventually attains a volume fraction of 90${\%}$ at 1.7~K and thus becomes the dominant state. Additionally, we obtain the direction of the internal magnetic field at the muon site by evaluating the data from all four positron detectors surrounding the sample: Forward-Backward (F-B) and Up-Down (U-D) (see Fig.~4e-f) (details are given in the Supplementary Information). The measured amplitude $F_{\rm UD}$ of the oscillations detected on U-D detectors shows the maximum amplitude (nearly 100 ${\%}$) in the temperature range between 1.7~K and 310~K and no oscillations are found on F-B detectors. This indicates that the static internal field is pointing towards the $c$-axis and no spin reorientation takes place down to 1.7~K. Above 310~K, the magnitude of $F_{\rm UD}$ is reduced by a factor of two, which indicates that the internal field is pointing somewhere in the $ab$-plane, which is consistent with the spin reorientation transition from $c$-axis to the $ab$-plane.

We note one important aspect. If the magnetic order would be fully static with the internal field pointing along the $c$-axis, only a weak depolarization of the ${\mu}$SR signal would be observed in the F-B detector. In contrast, a fast depolarization of the implanted muons is seen in a wide temperature range (see Fig.~4c). The fast depolarization of the ${\mu}$SR signal in the F-B detector is direct evidence for the involvement of fluctuations in the magnetic state of TbMn$_{6}$Sn$_{6}$ in the time window of the technique ( 10$^{-5}$~to~10$^{-12}$~s). That the fluctuations are the cause of the observed muon spin depolarization is also supported by the measurement under 300~mT applied in the F-B direction (longitudinal field geometry), which show a negligible field effect on the depolarization rate at 300~K (see the Supplementary Information). As seen in Fig.~4c, the depolarization rate ${\lambda}_{L}$ of the ${\mu}$SR signal in the F-B detectors shows a weak temperature dependence down to $T_{\rm C1}^{*}$~${\simeq}$~120~K, below which ${\lambda}_{L2}$ shows a large increase. It reaches its maximum at $T_{\rm {C1}}$~${\simeq}$~20~K and decreases rapidly at lower temperatures. The amplitude of the dynamic signal also decreases below 20~K, but it still exists even at the base-$T$ of 1.7~K. The increase of ${\lambda}_{L2}$ below $T_{\rm C1}^{*}$ is the signature of a slowing of magnetic fluctuations until a quasi-static order is established below $T_{\rm {C1}}$ and a coherent precession with slightly higher frequency than the original is seen. Therefore, the $T_{\rm {C1}}$ should be considered as a crossover temperature signaling a slowing down of magnetic fluctuations rather than being regarded as a true phase transition temperature.  Moreover, even at the base-$T$  both static patches (which correspond to the higher frequency) and dynamic patches (which correspond to lower frequency) coexist. Using the moment sizes determined from neutron experiments and the $c$-axis ferrimagnetic structure, muon stopping site and local field calculations (see Methods for more details) find a dipolar field of [0~0~1.41]~T at the only stable muon stopping site, which is in good agreement with the high-frequency Component~I. Any deviation from this $c$-axis-aligned structure will therefore produce a lower internal field; thus, muon stopping site calculations show that the structure within the static patches below 20~K is perfectly $c$-axis aligned. For fluctuations faster than the ${\mu}$SR time window, the technique probes their time-averaged structure. Therefore, the combined observations of both static coherent oscillations and magnetic fluctuations between 20 K and 310 K can be best understood if the time averaged (ferrimagnetic) structure has a net static internal field aligned with the $c$-axis, but with smaller time-averaged $c$-axis moment. This explains slightly lower (by 15~${\%}$) internal field at the muon stopping site created by fluctuating patches.

%%%%%%%%%%%%%%%%%%%%%%%%%%%%%%%%%%%%%%%%%%%%%%%%%%%%%%%%%%
\begin{figure*}[t!]
\includegraphics[width=1.0\linewidth]{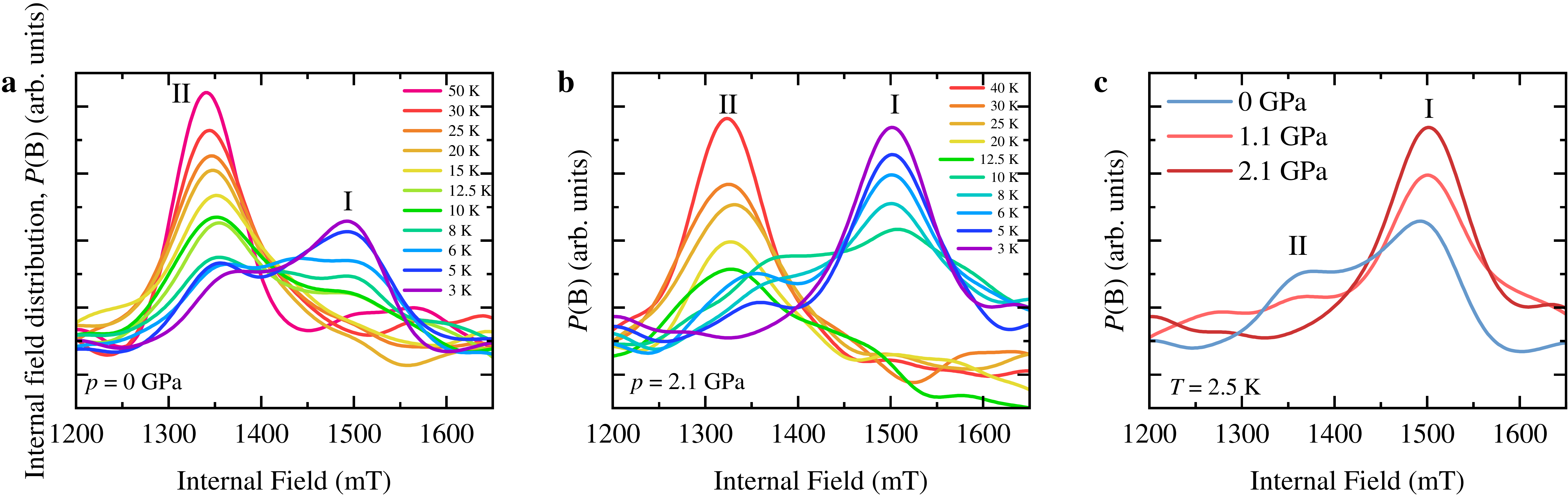}
\vspace{-0.5cm}
\caption{ \textbf{Hydrostatic pressure promoting the static out-of-plane ferrimagnetic state in TbMn$_{6}$Sn$_{6}$.}
Measured internal field distribution $P(B)$ recorded at various temperatures for the ambient pressure $p$ = 0 GPa (a) and 
maximum applied pressure of $p$ = 2.1 GPa (b). (c) $P(B)$ measured at the base-$T$ of 2.5 K at various pressures.}
\label{fig5}
\end{figure*}
%%%%%%%%%%%%%%%%%%%%%%%%%%%%%%%%%%%%%%%%%%%%%%%%%%%%%%%%%%%%%%%

%%%%%%%%%%%%%%%%%%%%%%%%%%%%%%%%%%%%%%%%%%%%%%%%%%%%%%%%%%
\begin{figure*}[t!]
\includegraphics[width=0.7\linewidth]{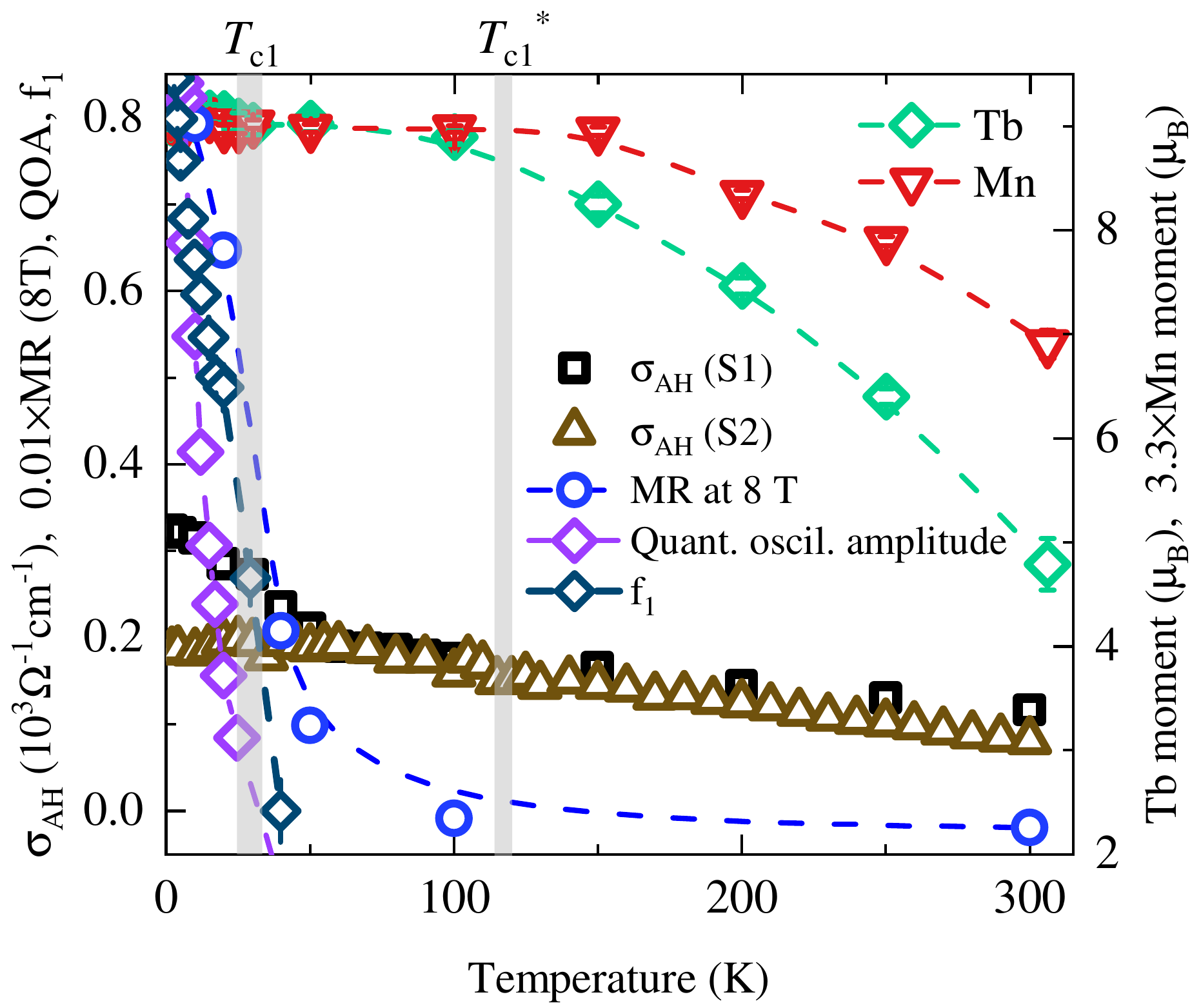}
\vspace{0cm}
\caption{ \textbf{Comparison between electronic and magnetic properties of TbMn$_{6}$Sn$_{6}$.}
The temperature dependence of the in-plane anomalous Hall conductivity, the relative fraction $f_{\rm 1}$ of the low-$T$ novel magnetic component,  the value of the magnetoresistance at 8 T (after Ref. \cite{Shuang}), the amplitude of the quantum oscillations (after Ref. \cite{TbNature}), and the magnetic moments of Tb as well as Mn, for TbMn$_{6}$Sn$_{6}$. $S_{1}$, and $S_{2}$ denote two different samples.}
\label{fig5}
\end{figure*}
%%%%%%%%%%%%%%%%%%%%%%%%%%%%%%%%%%%%%%%%%%%%%%%%%%%%%%%%%%%%%%%

\subsection{Hydrostatic pressure $\mu$SR measurements}

For further insight into the magnetic order and the novel low-temperature magnetic crossover in TbMn$_{6}$Sn$_{6}$, ZF-$\mu$SR experiments were carried out as a function of hydrostatic pressure. The probability distributions of internal fields sensed by the muon ensemble measured at $p$~=~0~GPa and 2.1~GPa
for various temperatures are shown in Figures~5a and b, respectively. Figure~6c shows the field distribution measured at base-$T$ of 2.5~K as a function of pressure. Remarkably, the fraction of the higher frequency component I, arising from patches with the static $c$-axis order, increases with pressure and eventually attains full volume fraction at 2.1~GPa. This implies that increased pressure stabilises the static out-of-plane ferrimagnetic ground state. From temperature dependent measurements, we also find that under pressure of 2.1~GPa, the onset temperature below which component I appears is higher by 5~K in comparison to ambient pressure.  These findings show the effective and volume wise tuning the competition between static and dynamic magnetic patches in this material with pressure.

\subsection{Discussion}

The combination of powder and single crystal neutron diffraction establishes the $c$-axis ferrimagnetic $P6/mm'm'$ structure in TbMn$_{6}$Sn$_{6}$ below $T_{\rm C2}$~=~310~K. ${\mu}$SR and macroscopic magnetization measurements uncover transition in TbMn$_{6}$Sn$_{6}$ at $T_{\rm C1}^{*}$~${\simeq}$~120~K. ${\mu}$SR shows that the transition at $T_{\rm C1}^{*}$ is related to a slowing down of magnetic fluctuations and is considered a crossover rather than a true phase transition. These magnetic fluctuations only become static below $T_{\rm C1}$~${\simeq}$~20~K, as seen by the additional precession frequency in the ${\mu}$SR signal. Considering the fact that neutron diffraction depicts the static order in the whole temperature range, we conclude that the moments fluctuate at a rate of the order of MHz, yet slower than the nearly instantaneous (ns-ps) time scale of neutron scattering. As ${\mu}$SR probes the time averaged structure for fluctuations faster than the time window, the combination of static coherent oscillations and magnetic fluctuations in the ${\mu}$SR signal can be understood if the time averaged structure at high temperatures is such that the net static internal field points along the $c$-axis. Muon stoping site calculations and local field analysis show that component I arises from the static ideal $c$-axis order and component II stems from dynamic fluctuations with smaller time averaged $c$-axis moment. The fact that the fraction of the Component I increases at the cost of Component II with decreasing temperature indicates that the slowing down process proceeds in the following fashion: when the fluctuations slow down and reach static enough conditions that we start observing the coherently precessing ${\mu}$SR signal below $T_{\rm C1}$~${\simeq}$~20~K, the patches of such static regions increase with decreasing temperature and the fraction of the corresponding signal increases. However, even at the base-$T$ of 1.7~K, Component~I does not acquire the full fraction and the moments are not yet fully static. This suggests that the establishment of a full volume static $c$-axis-aligned ferrimagnetic $P6/mm'm'$ order might appear only in the zero-temperature limit. The application of hydrostatic pressure supports this ideal $c$-axis-aligned phase, allowing for us to achieve a 100\% volume fraction at finite temperatures with the application of moderate pressure. The presence of magnetic fluctuations can also explain the magnetization data. In the SQUID measurements, which is a slower probe than ${\mu}$SR, a large difference between the FC and ZFC response is observed below $T_{\rm C1}^{*}$~${\simeq}$~120~K, e.g., in the region where fluctuations slow down. Since domains become more static, different out-of-plane ferrimagnetic domains tend to cancel out (anti-align) after ZFC. The domains try to minimize the total magnetization. However, if we FC even in low fields, then the domains align with the external fields instead. At temperatures higher than $T_{\rm C1}^{*}$~${\sim}$~120~K, the domains are highly dynamic on microsecond time scales and the overall effect is averaged out and no hysteresis is registered in the magnetization data.

In order to explore the correlation between the magnetic properties found in this work and the topological aspects \cite{TbNature,Shuang} of TbMn$_{6}$Sn$_{6}$, in Fig. 6 we compare the temperature dependence of the AHC, the MR and the amplitude of the quantum oscillations to the temperature evolution of the Tb and Mn moments as well as the amplitude of the low-$T$ magnetic component:

(1) We find that the amplitude of the quantum oscillations \cite{TbNature}, which reveals the Landau quantization from the bulk crystal, sets in below $T_{\rm C1}$~${\simeq}$~20~K when the ideal out-of-plane quasi-static magnetic component starts to appear (see Figure 6).

(2) It was previously  \cite{Shuang} shown that the MR of TbMn$_{6}$Sn$_{6}$ exhibits a quasi-linear field dependence (${\propto}$~$H^{1.1}$) only at temperatures below $T_{\rm C1}$~${\simeq}$~20~K. It was explained to stem from the quantum limit \cite{Abrikosov} of the linearly dispersive spin-polarized Dirac band with a large Chern gap. In Fig. 6, we point out that the MR changes sign from negative to positive across the magnetic crossover $T_{\rm C1}^{*}$~${\simeq}$~120~K and the absolute value of MR at 8~T increases substantially across $T_{\rm C1}$. Thus, from Fig. 5 it is evident that the Dirac band related MR effect is enhanced upon approaching the quasi static low-$T$ out-of-plane ferrimagnetic state.

(3) We find that the temperature dependence of ${\sigma}_{AH}(T)$ is strongly influenced by the low temperature magnetic crossover. Namely, while ${\sigma}_{AH}(T)$ for the two samples perfectly coincide with each other above $T_{\rm C1}^{*}$~${\sim}$~120~K, distinct temperature behavior is seen at lower temperatures. For one sample, ${\sigma}_{AH}$ increases with decreasing temperature with a stronger increase across $T_{\rm C1}$. For the second sample, the linear increase of ${\sigma}_{AH}$ over a wide temperature range changes into a nearly temperature independent behavior followed by a weak downturn across the temperature $T_{\rm C1}$. We also show that the ${\sigma}_{AH}(T)$ does not follow the temperature evolution of the Tb and Mn moments which indicates the intrinsic nature of the AHC in TbMn$_{6}$Sn$_{6}$, arising from the non-trivial magnetic-topological state.
Distinct behavior of the low temperature ${\sigma}_{AH}(T)$ in different samples might be caused by distinct magnetic domain distribution (different strains due to different cooling rate together with strong magnetostriction, may influence the magnetic domain distribution), which opens new opportunities for tunable topological magnetic states. This might also motivate a future direction of research to study the magneto-structural coupling in this topological system TbMn$_{6}$Sn$_{6}$.

It was shown previously \cite{TbNature} that TbMn$_{6}$Sn$_{6}$ exhibits spin-polarized Dirac fermions with a Chern gap and topological edge states within the Chern energy gap below ${\sim}$ 20 K. Using the fundamental magnetic kagome model, it was demonstrated that the topological gap is opened by combination of spin-orbit coupling and out-of-plane ferrimagnetic ordering in the kagome lattice and effectively the spinless Haldane model is realised by generating Chern gapped topological fermions. Our presented experimental results show that the onset of the topological electronic properties tied to the Dirac band is promoted only by true static out-of-plane ferrimagnetic order in TbMn$_{6}$Sn$_{6}$ and that the topological response is washed out by the slow magnetic fluctuations above $T_{\rm C1}$. The dynamics of the electrons are much faster than the time scale of these slow magnetic fluctuations. Therefore, one can think of the electrons moving in the "static" random field created by the domain distributions that affects the coupling of the Dirac electrons to the local magnetic structure and the effective topological Chern gap. Moreover, there may be parasitic conduction channels created by the slowly fluctuating domain walls etc. that affect transport. Furthermore, the slow fluctuations and broad transition from $T_{\rm C1}^{*}$ to $T_{\rm C1}$ may involve more of the low energy Dirac fermions, such that they may have a spin component in the plane that do not open a topological mass gap solely by in-plane spin-orbit coupling. In fact, even below $T_{\rm C1}$, we would expect magnetic fluctuations from the ordered ferrimagnet (such as spin waves etc.), that would still have an effect in reducing the topological response, but the emergence of dense ordered static patches certainly help protect these from the slow dynamical fluctuations of the domains. Since we show the presence of out-of-plane ferrimagnetic structure (comprised by Tb and Mn moments) with magnetic fluctuations below 320K, the exciting perspective arises of a magnetic system in which the topological response can be obtained at room temperature just by switching slow fluctuations into a static state e.g., without substantial modification of magnetic structure.

\section{Summary}

The exploration of topological electronic phases that result from strong electronic correlations is a frontier in condensed matter physics. Kagome lattice systems are an ideal setting in which strongly correlated topological electronic states may emerge. Our key finding is the identification of a novel low-temperature magnetic crossover in TbMn$_{6}$Sn$_{6}$, which occurs concurrently with its topological properties. Namely, a combination of neutron diffraction, ${\mu}$SR and magnetization measurements indicate that the system TbMn$_{6}$Sn$_{6}$ exhibits out-of-plane ferrimagnetic $P6/mm'm'$ order between Mn and Tb moments with slow magnetic fluctuations in a wide temperature range 1.7~K~--~315~K. The moments fluctuate at a rate on the order of MHz, yet slower than the nearly instantaneous time window of neutron scattering. The fluctuations start to slow down below $T_{\rm C1}^{*}$~${\simeq}$~120~K, forming quasi-static patches only below $T_{\rm C1}$~${\simeq}$~20~K and become enhanced in a volume-wise manner upon further lowering of temperature. The established quasi-static magnetic state below $T_{\rm C1}$ takes the ideal out-of-plane ferrimagnetic structure. Remarkably, the application of pressure increases the volume fraction of such patches and pressure of 2.1 GPa stabilises static out-of-plane ferrimagnetic ground state in the whole volume of the sample.
% differing slightly from the} \textcolor{red}{slowly fluctuating, likely canted ferrimagnetic structure} \textcolor{red}{present from $T_{\rm C2}$ to $T_{\rm C1}$ where all moments contribute to a net $c$-axis ferrimagnetism and exhibit a zero net in-plane moment.} 
The interplay between this intricate magnetism and the spin-orbit coupled band structure further induces non-trivial variations of its topological properties, which is characterized by the influence of the low temperature magnetic crossover on the temperature evolution of AHC as well as by the appearance of the large quasi-linear MR and the quantum oscillations, related to the quantized Landau fan structure featuring spin-polarized Dirac dispersion with a large Chern gap, within the low-$T$ magnetic state. These experiments indicate that observed slow magnetic fluctuations above $T_{\rm C1}$~${\simeq}$~20~K are detrimental to the topological Chern gap and will stimulate further theoretical studies to obtain a microscopic understanding of the relation between the slow fluctuations, low-temperature volume-wise magnetic evolution of the static $c$-axis ferrimagnetic patches and the quantum-limit Chern gapped phase.

\section{METHODS}

\textbf{General remarks}: We concentrate on the high resolution \cite{GPSamato,GuguchiaMoTe2} muon spin relaxation/rotation (${\mu}$SR) measurements of the temperature dependence of the magnetic moment as well as on the magnetically ordered volume fraction in the single crystals of TbMn$_{6}$Sn$_{6}$ and on the high resolution neutron powder diffraction of the magnetic structure at various temperatures. In a ${\mu}$SR experiment, positive muons implanted into a sample serve as an extremely sensitive local probe to detect small internal magnetic fields, ordered magnetic volume fractions and magnetic fluctuations in the bulk of magnetic materials  \cite{Dalmas}. Density functional theory calculations were used to explore the electronic band structure and to calculate the Berry curvature-induced anomalous Hall conductivity. Neutron diffraction is used to determine the magnetic structure. The techniques of ${\mu}$SR, neutron diffraction, transport and DFT complement each other ideally as we are able to study the detailed temperature dependence of the magnetic order parameter and ordered volume fractions with ${\mu}$SR experiments, and correlate them with the measured and calculated anomalous Hall conductivity.\\

\textbf{Sample preparation}: Details of the sample preparation, characterization, and the Hall effect measurements are described elsewhere \cite{TbNature,Shuang}.\\

\textbf{High resolution neutron powder diffraction measurement}:  The magnetic and crystal structure of TbMn$_{6}$Sn$_{6}$ have been studied with neutron powder diffraction (NPD) experiments carried out at the Swiss Neutron Spallation Source (SINQ), at the Paul Scherrer Institute in Villigen, Switzerland. Several single crystal samples were crushed into a fine powder and loaded into a 6~mm diameter vanadium can. The diffraction patterns were collected on the High Resolution Powder diffractometer for Thermal neutrons (HRPT) using wavelengths $\lambda$~=~1.8857~{\AA} and $\lambda$~=~2.449~{\AA} (Ge-monochromator, 2$\theta_{max}$~=~$160^{\circ}$, 2$\theta_{step}$~=~$0.1^{\circ}$) in the temperature range from 1.50-306~K. High-statistic acquisitions for magnetic structure refinements were made.\\

\textbf{Temperature and field dependent single crystal neutron diffraction measurements}:
Single crystal neutron diffraction was performed on our sample using the thermal single crystal diffractometer ZEBRA, at the SINQ/PSI. A single hexagonal crystal of approximately 4~mm diameter and 0.4~mm thickness, the $c$-axis was along the thin dimension of the crystal. After the single-crystallinity check and alignment by Laue X-ray diffraction, the sample was mounted to a 0.3-mm-thick rectangular sheet of aluminum with GE varnish, then bound with thin aluminum wires before being wrapped in 0.1~mm aluminum foil. Two different setups were used -  with a cooling machine or furnace on a Eulerian cradle and with a cryomagnet with vertical field up to 10~T. For both experiments the incident neutron wavelength of 1.18~{\AA} was selected by the Ge-monochromator. Datasets at four temperatures - 5~K, 150~K, 300~K and 425~K - were measured on the  Eulerian cradle. In the magnet experiment the magnetic field was applied along the (1-10) direction, within the $ab$-plane.
The datasets at three temperatures - 2K, 70K and 250K - were collected.
Field dependence was measured up to 8~T at 2~K for the (110), (002) reflections and at 230~K for the (001), (002), and (110) reflections, respectively. From our collected datasets, we performed refinements with the FullProf suite using the ISODISTORT program online to generate the Shubnikov subgroups. The longer wavelength of $\lambda$~=~2.3~{\AA} (PG monochromator) and the 2D detector were used to detect any incommensurate feature near or splitting of the 002 reflection on cooling between 225~K and 25~K. No new features have been found.\\

\textbf{${\mu}$SR experiment}: In a ${\mu}$SR experiment nearly 100 ${\%}$ spin-polarized muons (${\mu}$$^{+}$)
are implanted into the sample one at a time. The positively
charged ${\mu}$$^{+}$ thermalize at interstitial lattice sites, where they
act as magnetic microprobes. In a magnetic material the
muon spin precesses in the local field $B_{\rm \mu}$ at the
muon site with the Larmor frequency 2${\pi}$${\nu}_{\rm \mu}$ = $\gamma_{\rm \mu}$/(2${\pi})$$B_{\rm \mu}$ [muon
gyromagnetic ratio $\gamma_{\mu}$/(2${\pi}$) = 135.5 MHz T$^{-1}$].

Longitudinal-field (LF) and zero-field ${\mu}$SR experiments on the single crystalline samples of TbMn$_{6}$Sn$_{6}$ were performed at the ${\pi}$M3 beamline of the Paul Scherrer Institute (Villigen, Switzerland), using the low background GPS instrument \cite{GPSamato}. The specimen was mounted in a He gas-flow cryostat and CCR with the $c$-axis parallel to the muon beam direction in order to cover the temperature range between 1.7 K and 400 K.

Hydrostatic pressure $\mu$SR experiments were performed at the $\mu$E1 beamline of the Paul Scherrer Institute using the General Purpose Decay-Channel Spectrometer (GPD) instrument \cite{GuguchiaPressure}. An ensemble of many single crystals were compacted into a MP-35N pressure cell, mounted in a He gas-flow cryostat able to achieve a base temperature of 2.5~K. \\

\textbf{Analysis of ZF-${\mu}$SR data}: The ZF-${\mu}$SR spectra from Up-Down positron detectors were fitted using the following model \cite{AndreasSuter}:

	\begin{widetext}
		\begin{equation}
		\label{eq:ZFPolarizationFit1}
		A_{\textrm{ZF,UD}}(t) = \sum_{j = 1}^{3}  \left( f_{T,j} \cos(2\pi\nu_j t + \phi) e^{-\lambda_{T,j} t} \right) +
		f_{L,j} e^{-\lambda_{L,j} t}.
		\end{equation}
	\end{widetext}
	The model (1) is an anisotropic magnetic contribution
	characterized by an oscillating transverse component and a slowly relaxing longitudinal
	component.  The longitudinal component arises due to the parallel orientation of the muon spin
	polarization and local magnetic field components. In polycrystalline samples with therefore randomly oriented fields
	this results in a so-called one-third tail with $f_{\rm L} = \frac{1}{3}$. For single crystals, $f_{\rm L}$
	varies between unity and zero as the orientation between field and polarization changes from being
	parallel to perpendicular. Note that the whole volume of the sample is magnetically ordered in the whole investigated temperature range. Depending on the temperature range, either a single, or two distinct well separated precession frequencies can be clearly seen in the ${\mu}$SR spectra. Spectra near the base-$T$ requires the addition of the third broad oscillating component with the internal field value very close to the one, observed at high temperatures. In addition, the spectra below 20 K are also characterised by a missing small fraction of the initial asymmetry, which points to a more disordered static state below 20 K than the one above.\\
	
	The ZF-${\mu}$SR spectra from Forward-Backward positron detectors were fitted using the following dynamic model:

	\begin{widetext}
		\begin{equation}
		\label{eq:ZFPolarizationFit2}
		A_{\textrm{ZF,FB}}(t) = A_{fast}e^{-\lambda_{FB,fast} t} + A_{slow}e^{-\lambda_{FB,slow}t}  .
		\end{equation}
	\end{widetext}
	
	The two terms represent a fast and a slow relaxation component, respectively, and $\lambda_{FB,fast}$ and
	$\lambda_{FB,slow}$ are the muon-spin-relaxation rates for each component.\\
	
\textbf{Calculation of the muon stopping site}: The DFT-based computer simulations carried out in this work were performed with the CASTEP \cite{DFT1} code. The crystal structure of TbMn$_{6}$Sn$_{6}$ used for the computation was obtained from the Inorganic Crystal Structure Database via the CrystalWorks portal. A plane wave cutoff of 800~eV for these calculations was chosen by converging energy and forces using the automated tool CASTEPconv \cite{DFT2}. As regards the $k$-point grid size, a high-density 12$\times$12$\times$6 Monkhorst-Pack $k$-point grid \cite{DFT3} was used. This produced forces accurate well within an error of 0.05~eV/\AA, which was used as the limit tolerance for geometry optimization. Geometry optimization on the structure was performed with a LBFGS algorithm, fixing the unit cell parameters to their experimental values, to a tolerance of 0.05 eV/\AA~for the forces. The LDA exchange-correlation functional was used in combination with auto-generated ultrasoft pseudopotentials.  The DFT calculations were spin-polarized calculations, with the quantization axis along the [001] direction and with initial magnetic moments of Tb~=~-9.0~${\mu}_{B}$ and Mn~=~2.4~${\mu}_{B}$, which are the experimental magnetization values at $T$~${\simeq}$~1.8~K.

The determination of the muon stopping sites was performed using the Unperturbed Electrostatic Method (UEP), as implemented in the software package pymuon-suite \cite{DFT4}, which provides various utilities to estimate the muon stopping sites \cite{DFT5, DFT6}.  The UEP method uses Density Functional Theory (DFT) calculations to estimate the host material's electrostatic potential plus a combination of mathematical analysis and clustering techniques to estimate potential muon stopping sites.  Single stable stopping site was identified. which are separated by potential barriers on the order of at least 0.2~eV, indicating that the muons are locked into these sites after thermalization.

Subsequent calculations of the internal dipolar fields at the potential muon sites were performed using the Python package muesr \cite{DFT7}.  A sphere with a radius large enough to encapsulate a 100$\times$100$\times$100 supercell was constructed and dipole summation over all moments in the sphere was performed to find the local field at the muon sites.  The magnetic structure considered was the experimentally proposed $c$-axis-aligned magnetic structure, with magnetic moments of Tb~=~-9.0~${\mu}_{B}$ and Mn~=~2.4~${\mu}_{B}$.  The local dipolar field for the muon stopping site indicated in the inset of Figure 4d is [0.00002~0.00001~1.40588]~T, which is in reasonable agreement with the value and direction of the experimentally observed local field for component $I$ at $T$~${\textless}$~10~K.\\

\textbf{Data availability}: All relevant data are available from the authors. The data can also be found at the following link http://musruser.psi.ch/cgi-bin/SearchDB.cgi.\\

%%%%%%%%%%%%%%%%%
\section{Acknowledgments}~
The ${\mu}$SR experiments were carried out at the Swiss Muon Source (S${\mu}$S) Paul Scherrer Insitute, Villigen, Switzerland using the high resolution GPS instrument (${\pi}$M3 beamline). The neutron diffraction experiments were performed at the Swiss spallation neutron source SINQ (HRPT and ZEBRA diffractometers), Paul Scherrer Institute, Villigen, Switzerland. The magnetization measurements were carried out on the MPMS device of the Laboratory for Multiscale Materials Experiments, Paul Scherrer Institute, Villigen, Switzerland. Z.G. and C.M. thank Romain Sibille and Dariusz Jakub Gawryluk for their useful discussions. T. N. and S. S. T. acknowledge support from the European Research Council (ERC) under the European Unions Horizon 2020 research and innovation programm (ERC-StG-Neupert757867-PARATOP). S. S. T. is additionally supported by the grant No. PP00P2 176877 by the Swiss National Science Foundation. X. L. was supported by the China Scholarship Council (CSC). This work was also supported by the Swiss National Science Foundation (R`Equip grant no. 206021${\_}$139082)\\
Z.Q.W. is supported by DOE grant No. DE-FG02-99ER45747.

%\textbf{\section{Author Contributions:}}
%Z.G. conceived the study. ${\mu}$SR experiments and corresponding data analysis: Z.G., C.M., and H.L.. Neutron diffraction experiments and analysis: C.M., O.Z., V.A.P., and Z.G.. Magnetization measurements: C.M., M.M., and Z.G.. STM experiments: J.-X.Y. and T.C. in consultation with M.Z.H. Growth of single crystals: W.M., and S.J.. Conductivity measurements: W.M., and S.J. Band structure and AHC calculations: X.L., S.S.T., and T.N.. Writing the paper: Z.G., and C.M. with contributions from all authors.
%All authors discussed the results, interpretation and conclusion.\\

\textbf{Competing interests:} The authors declare that they have no competing interests.\\

\newpage

\renewcommand{\figurename}{Supplementary Figure}

\renewcommand\thefigure{\arabic{figure}}
\setcounter{figure}{0}

\begin{center}
\textbf{Supplementary Information} \\
\end{center} 

 TbMn$_{6}$Sn$_{6}$ containing the heavy rare earth Tb crystallizes in a HfFe$_{6}$Ge$_{6}$-type structure (space group $P6$/$mmm$). It is composed of hexagonal Tb layers containing Sn atoms and Mn kagome nets, stacked in the sequence -Mn-Tb-Mn-Mn-Tb-Mn- along the $c$-axis (Figure S1a-b). Possible atomic assignment of the kagome lattice and a magnified image of scanning tunneling microscopy of the manganese-terminating surface is shown in Figures S1c-d.

\section{Temperature dependence of spin fluctuations and direction of the internal magnetic field}

 Apart from the magnetic ordering temperature and the magnetic fraction, it is instructive to evaluate the temperature dependence of the spin fluctuations, as well as the direction of the internal magnetic field. This was done through zero-field (ZF) ${\mu}$SR experiments. The measurements were performed with the initial muon spin polarization at an angle of $45^{\circ}$ with respect to the $c$-axis of the crystal, as illustrated in Fig.~S2h. The sample was surrounded by two pairs of detectors: Forward-Backward (F-B) and Left-Right (U-D). Figures~S2a, d and f show representative zero-field (ZF) ${\mu}$SR time spectra for the single crystal TbMn$_{6}$Sn$_{6}$, recorded using U-D at 1.8~K, 300~K, and 340~K, respectively. Figures~S2b, e and g show representative zero-field (ZF) ${\mu}$SR time spectra for the single crystal TbMn$_{6}$Sn$_{6}$, recorded using F-B at 1.8~K, 300~K, and 340~K, respectively.  In a single crystal with a fully magnetically ordered state, the amplitude of the oscillation is proportional to the angle between the muon spin polarization and the internal field ${\theta}$ (see Fig.~S2i, where the muon spin precession around the internal magnetic field, for two extreme cases $B_{\rm int}$ ${\perp}$ $c$ and $B_{\rm int}$ ${\parallel}$ $c$, is schematically illustrated). For the internal field direction, shown in the top panel of Fig.~S2i, the ${\mu}$SR signal from F-B detectors exhibits the maximum amplitude and no oscillations will be detected in the U-D detectors. The opposite will be observed for the configuration shown in the bottom panel of Fig.~S2i. Thus, by evaluating the data from all four detectors one can obtain useful information on the direction of the internal field and therefore the local spin configuration. At $T$ = 1.8~K and 300~K, the oscillations with maximum amplitude (see Fig.~S2a~and~d, oscillating around zero) are observed on U-D detectors, while no oscillations were found on F-B detectors. This indicates that the internal field is pointing towards the $c$-axis from 300~K down to 1.8~K and no spin reorientation takes place in this temperature range. We further note one important aspect; if the magnetic order is fully static with the internal field pointing along the $c$-axis, only a weak depolarization of the ${\mu}$SR signal would be observed on the F-B detector. In contrast, a fast depolarization of the implanted muons is seen in a wide temperature range (see Fig.~S2c). The fast depolarization of the ${\mu}$SR signal on F-B is direct evidence for the fluctuating magnetic moments in TbMn$_{6}$Sn$_{6}$. That the fluctuations are the origin of the observed muon spin depolarization is also supported by measurements under 300~mT, which shows negligible field effect on the depolarization rate at 300~K (see Figure~S2e). If the depolarization were caused by a wide distribution of static fields, then a small field would be enough to fully recover the muon spin polarization. Above 310~K, oscillations with equal amplitude are observed on U-D and F-B detectors. Accordingly, the magnitude of $F_{UD}$ is reduced by a factor of two, which indicates that the internal field is pointing somewhere in the $ab$-plane, consistent with the spin reorientation transition from $c$-axis to the $ab$-plane, reported previously from neutron diffraction studies \cite{Idrissi1991}.

\section{High resolution neutron powder diffraction measurement in TbMn$_{6}$Sn$_{6}$}
    The magnetic and crystal structure of TbMn$_{6}$Sn$_{6}$ have been studied via neutron powder diffraction (NPD) experiments carried out at the Swiss Neutron Spallation Source (SINQ, at the Paul Scherrer Institute in Villigen, Switzerland. Several single crystal samples were crushed into a fine powder and loaded into a 6~mm diameter vanadium can. High-statistic diffraction patterns were acquired on the High Resolution Powder diffractometer for Thermal neutrons (HRPT) using wavelengths $\lambda$ = 1.8857 {\AA} and $\lambda$ = 2.449 {\AA} (Ge [822], 2$\theta_{max}$ = $160{^\circ}$, 2$\theta_{step}$ = $0.1{^\circ}$) in the temperature range from 1.5-250~K. The structure was well refined by Rietveld refinements to the raw neutron diffraction data, employing a hexagonal lattice structure in the space group $P6/mmm$, $No.$~191. An example of the refinement profile at various temperatures is shown in Fig.~S4, indicating a good fit. Only a very small impurity phase has been detected, with a maximum peak intensity at 2$\theta$ = $37.6692^{\circ}$ at 1.5~K. This secondary phase has been identified as elemental tin, probably originating from residual flux on the surface of the crystals, similar to previous reports \cite{Clatterbuck1999}. High-statistic acquisitions were also performed with a wavelength $\lambda$~=~2.449~{\AA} to gain additional access to low-$q$ regions of the diffraction pattern and were unable to detect any additional peaks. Unfortunately, as the magnetic ordering temperature of 425~K \cite{Yin2020} was not achievable in the experimental setup, we could not observe the purely nuclear structure in the paramagnetic phase. Over the entire measured temperature range 1.5-250~K, no new peaks were observed, consistent with the ferrimagnetic $P6/mmm$, $No.$ 191 structure and previous reports. We also performed a low-statistic measurement at 306~K, the highest temperature achievable with our cryostat, and were able to refine the lattice parameters and magnetic moments, but were unable to achieve the resolution necessary to observe the presence of any new peaks (ie. the (0~0~1) peak) appearing due to canting from the imminent spin reorientation transition as reported in \cite{Idrissi1991}.

	Fits to this data were performed using the FullProf suite \cite{Fullprof1993}, with the use of the internal tables for neutron scattering lengths. Symmetry analysis was performed using the ISODISTORT tool based on ISOTROPY software \cite{{Stokes1988}, {Campbell2006}}. With ISODISTORT we determined the magnetic Shubnikov groups present within the maximal subgroup, and attempted fitting the collected patterns at 1.5~K, 5~K, and 250~K to many of the different maximal subgroups. In particular, we focused on comparing in- and out-of-plane structures, and models allowing different sites for the same atomic species. Specifically, we tested: all variations of the $P6/mmm$ hexagonal space group, with various models allowing either in-plane or out-of-plane magnetic moments; $Cm'm'm$, SG~No.~65.485, allowing two different Mn sites with $m_{z}$ components; $Cmm'm'$ SG~No.~65.486, allowing $m_{x}$ and $m_{y}$ in-plane magnetic moments; and finally $P2'/m$ and $P2'/m'$, SG's~No.~10.44 and 10.46, respectively, which allow $m_{x}$ and $m_{z}$ magnetic components. For powder diffraction data, only two structures achieved good fits to the collected data: $P6/mm'm'$ No.~191.240 and $Cm'm'm$ SG~No.~65.485. $P6/mm'm'$ consists of two sublattices oriented antiferromagnetically relative to one another, as shown in Figure~2a. The Tb and Mn magnetic moments are collinear, and point along the $z$-direction. The moments of each sublattice are parallel among themselves and antiparallel to the other sublattice. Thus, the overall structure is a collinear ferrimagnetic  structure. The model with $Cm'm'm$ symmetry is also $c$-axis ferrimagnetic, essentially the same as $P6/mm'm'$ but allowing two different Mn sites. Of these, the higher-symmetry $P6/mm'm'$ was selected as the preferred candidate structure, because the refinements in $Cm'm'm$ converged to two very similar magnetic moment sizes on the two Mn sites (nearly the same within the error bar). We note that the model $P6/mm'm'$ describes the data equally well over the whole temperature range, consistent with previous reports \cite{Idrissi1991}. With this structure we clearly see the segregated kagome layers containing purely Mn atoms, separated by hexagonal layers of Sn and Tb \cite{Yin2020}.

\section{Temperature and field dependent single crystal neutron diffraction measurements in TbMn$_{6}$Sn$_{6}$}
     Single crystal neutron diffraction was performed on our sample using ZEBRA, at the SINQ/PSI. A single crystal sheet was selected that was approximately 4~mm in diameter and 0.4~mm in thickness. The flux-grown crystals were small, thin sheets; the thin axis was determined to be along the crystallographic $c$-axis (see Fig.~S6). The selected crystal had a semi-hexagonal shape; when aligned using Laue X-ray diffraction, it was very clear that the $c$-axis was along the thin part of the crystal, and the $a-$ and $b-$axes were aligned with the hexagonal edges of the crystal (see Fig.~S6). After the alignment and single-crystallinity of the selected sample were checked by Laue X-ray diffraction, the sample was mounted to a 0.3~mm-thick rectangular sheet of aluminum with GE varnish, then bound with thin aluminum wires before being wrapped in 0.1~mm aluminum foil.

The sample was then mounted in a cryomagnet with vertical field up to 10 Tesla or an Eulerian cradle. State-of-the-art Small Angle Neutron Scattering (SANS) measurements were also performed at the SANS-II instrument at SINQ, PSI, on a 143~mg single crystal of TbMn$_{6}$Sn$_{6}$ in a $q$-range from 0.25~\AA$^{-1}$ to 0.005~\AA$^{-1}$. Over the entire measured temperature- and $q$-range, no modulation of the magnetic structure which is incommensurate was determined by SANS, in both in- and out-of-plane orientations. Full matrices of reflections were measured with the point detector at 435~K, 250~K, 70~K, and 2~K, and in 5~T applied magnetic field at 250~K. The magnetic field was applied along the $ab$-plane. Field dependence was measured up to 8~T at 2~K for the (110), (002) reflections and at 230~K for the (001), (002), and (110) reflections, respectively. At 230~K, we found a spin reorientation transition occurring at H$_{c}$~=~4.8~T applied field, marked by the sudden appearance of diffraction intensity at the (0~0~1) peak position (see Fig.~S5). At 2~K, there was no visible transition up to 10~T applied field. 
	
The data was first refined in the paramagnetic state, and then the crystallographic sites and temperature factors were fixed and the data in the magnetically ordered states were refined. The full matrix of diffraction intensities collected within the paramagnetic regime at 435~K was refined in the $P6/mmm$ structure. The successful refinement and goodness of fit serve as a confirmation of the single crystallinity and quality of our measured sample.

 From our collected matrix of diffraction intensities and points, we performed refinements with the FullProf suite using the ISODISTORT program online to generate the Shubnikov subgroups. An ideal fit to the collected diffraction intensities was obtained at 250~K using the same magnetic structure used for the NPD refinements, $P6/mm'm'$, space group $No.$ 191.240 (see Fig.~S6b). The data was fitted to the same magnetic structure found via NPD refinements ($P6/mm'm'$, space group $No.$ 191.240, see Fig.~S6b). The ideal out-of-plane ferrimagnetic $c$-axis aligned $P6/mm'm'$ structure has the best fit for the data collected at 150 K and 300 K;  however, at the lowest temperature, 5~K, the out-of-plane ferrimagnetic $P2/m$, No.~10.42 achieved a slightly better fit to the data. This lower-symmetry structure allows for three different Mn sites and a mixing of two irreducible representations. Both $P6/mm'm'$ and $P2/m$ are characterised by a perfectly $c$-axis aligned structure. Thus, the ideal out-of-plane hexagonal $P6/mm'm'$ structure conclusively fits best to the collected neutron diffraction data.\\  
  	
\subsection{Transport measurements}

Extended Hall conductivity and magnetoresistance measurements on TbMn$_{6}$Sn$_{6}$ were recently reported \cite{Shuang,TbNature}. Here, we show the AHC for two pieces of the single crystals from the same batch of TbMn$_{6}$Sn$_{6}$, which was obtained using the formula ${\sigma}_{AH}$ = ${\rho}_{AH}$/${\rho}_{xx}^{2}$, where  ${\rho}_{AH}$ and ${\rho}_{xx}$ are the anomalous Hall and the zero-field longitudinal resistivity, respectively. 
When plotting ${\rho}_{AH}$ against the square of the longitudinal resistivity ${\rho}_{xx}^{2}$, a linear scaling is observed \cite{TbNature}, which by definition indicates that ${\rho}_{AH}$ is dominated by an intrinsic contribution arising from the Berry curvature fields in total anomalous terms. From the linear fit, the intrinsic contribution was estimated. 
It is noteworthy that  for one sample ${\sigma}_{AH}$  shows an upturn across $T_{\rm {C1}}$ ${\simeq}$ 20 K, while for the other sample from the same batch ${\sigma}_{AH}$ shows a weak downturn at low temperatures. However, the downturn is not as sharp as upturn. These seemingly inconsistent results and previous transport measurements will be discussed below in connection with the magnetic properties.

%%%%%%%%%%%%%%%%%%%%%%%%%%%%%%%%%%%%%%%%%%%%%%%%%%%%%%%%%%
\begin{figure*}[t!]
\includegraphics[width=0.9\linewidth]{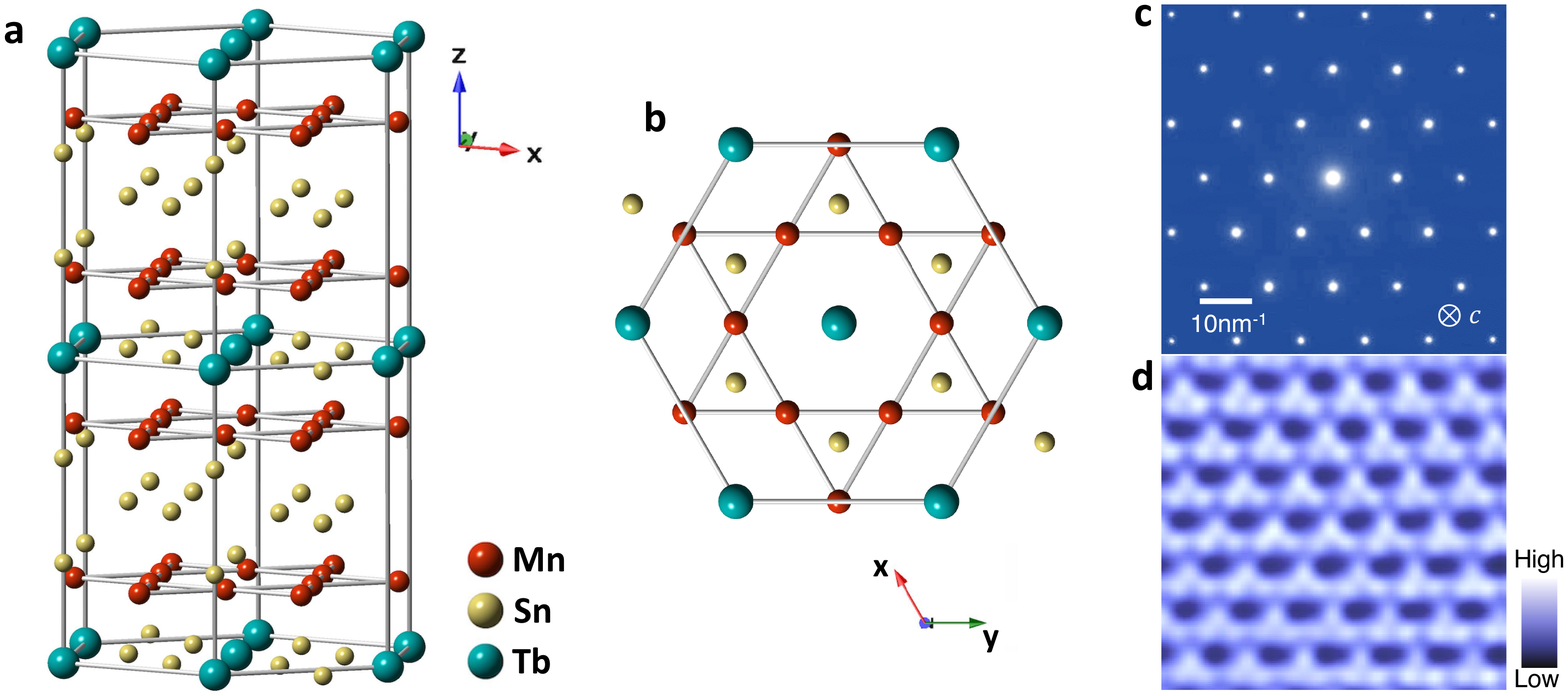}
\vspace{-3.7cm}
\caption{ \textbf{Crystal structure of TbMn$_{6}$Sn$_{6}$.}
Three dimensional representation (a) and top view (b) of the atomic structure of TbMn$_{6}$Sn$_{6}$.
The Mn atoms construct a Kagome lattice (red middle size circles), while the Tb (brown small size circles) and Tb atoms (dark green large size circles) form a honeycomb and a triangle structure, respectively.
(b) Kagome lattice structure of the Co$_{3}$Sn layer. 
(c) Possible atomic assignment of the kagome lattice. (d) Magnified image of scanning tunneling microscopy of the manganese-terminating surface taken at 4.2 K. Data are taken at the tunneling junction: $V$ = 50 mV, $I$ = 0.8 nA, $T$ = 4.2 K.}
\label{fig1}
\end{figure*}
%%%%%%%%%%%%%%%%%%%%%%%%%%%%%%%%%%%%%%%%%%%%%%%%%%%%%%%%%%

%%%%%%%%%%%%%%%%%%%%%%%%%%%%%%%%%%%%%%%%%%%%%%%%%%%%%%%%%%
\begin{figure*}[t!]
\includegraphics[width=1.0\linewidth]{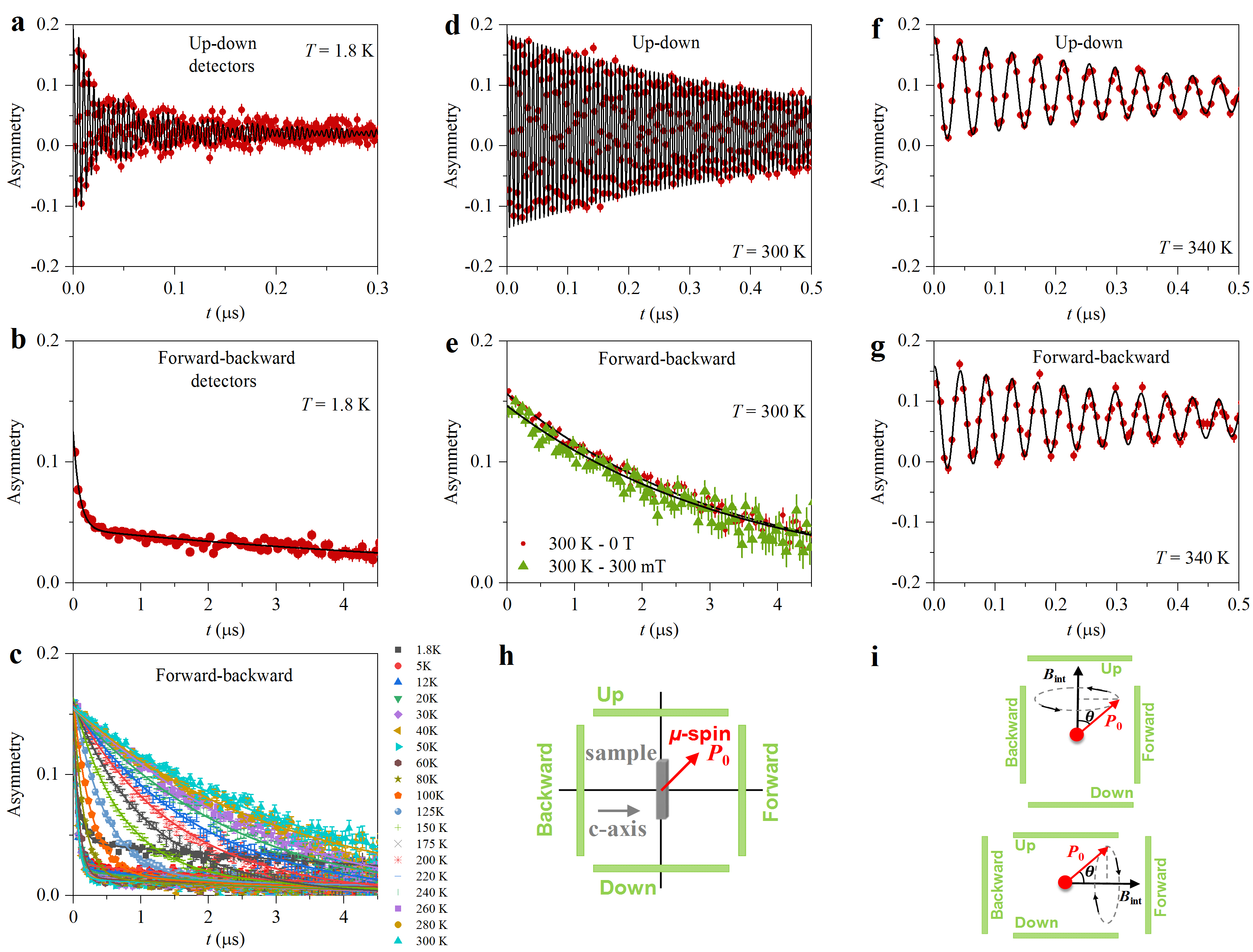}
\vspace{-0.0cm}
\caption{ \textbf{Zero-field ${\mu}$SR signals for TbMn$_{6}$Sn$_{6}$.}
Zero-field ${\mu}$SR signals for TbMn$_{6}$Sn$_{6}$ recorded at 1.8 K, 300 K and 340 K, for up-down (a,d,f) and forward-backward detectors (b,e,g). (c) Zero-field ${\mu}$SR signals from F-B detectors, recorded at various temperatures. (h) A schematic overview of the experimental setup for the muon spin forming $45^{\circ}$ with respect to the $c$-axis of the crystal. The sample was surrounded by four detectors: Forward (F), Backward (B), Up (U) and Down (D). (i) Schematic illustration of the muon spin precession around the internal magnetic field for two cases: (top) The field is perpendicular to the $c$-axis and points towards the L-detector. ${\theta}$ is the angle between the magnetic field and the muon spin polarization at $t$ = 0. (bottom) The field is parallel to the $c$-axis of the crystal and points towards the F-detector.}
\label{fig1}
\end{figure*}
%%%%%%%%%%%%%%%%%%%%%%%%%%%%%%%%%%%%%%%%%%%%%%%%%%%%%%%%%%

%%%%%%%%%%%%%%%%%%%%%%%%%%%%%%%%%%%%%%%%%%%%%%%%%%%%%%%%%%%%%%
\begin{figure*}[t!]
\centering
\includegraphics[width=1.0\linewidth]{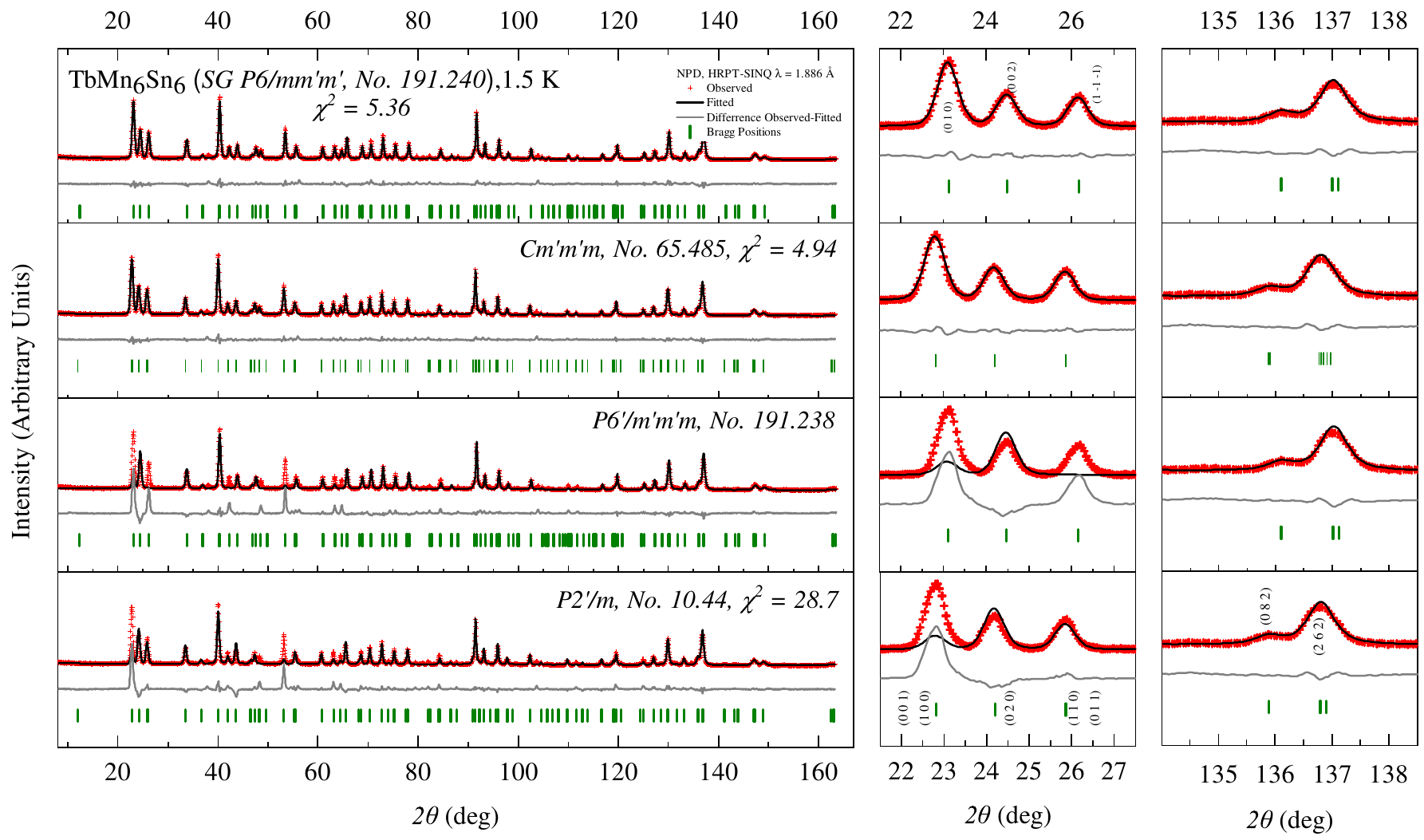}
\vspace{-0.8cm}
\caption{ (Color online) \textbf{Magnetic structure comparison for TbMn$_{6}$Sn$_{6}$.} 
In order to solve the magnetic structure, we performed refinements on the NPD patterns with different structural and magnetic models. Possible magnetic Shubnikov groups were generated from the hexagonal $P6/mmm$ space group using ISODISTORT \cite{{Stokes1988}, {Campbell2006}}. The maximal symmetry subgroup is $P6/mm'm'$, $No.$ 191.240, which consists of Tb and Mn sublattices oriented antiferromagnetically relative to one another, with moments along the $c$-axis. The other model which fits comparably well ($Cm'm'm$, $No.$ 65.485) has lower symmetry, allowing for two independent Mn sites and moments. Space group $No.$ 191.238 is the isostructural model to 191.240, however with magnetic moments confined to the $ab$-plane. The final structure considered here was incorporated because it is a low-symmetry monoclinic model, $P2'/m$, SG $No.$ $10.44$ which allows for different Mn and Tb sites and has no structural constraints. With this model, $mx$ and $mz$ moments are allowed. 
}
\label{fig2}
\end{figure*}
%%%%%%%%%%%%%%%%%%%%%%%%%%%%%%%%%%%%%%%%%%%%%%%%%%%%%%%%%%%%%%

%%%%%%%%%%%%%%%%%%%%%%%%%%%%%%%%%%%%%%%%%%%%%%%%%%%%%%%%%%%%%%
\begin{figure*}[t!]
\centering
\includegraphics[width=1.0\linewidth]{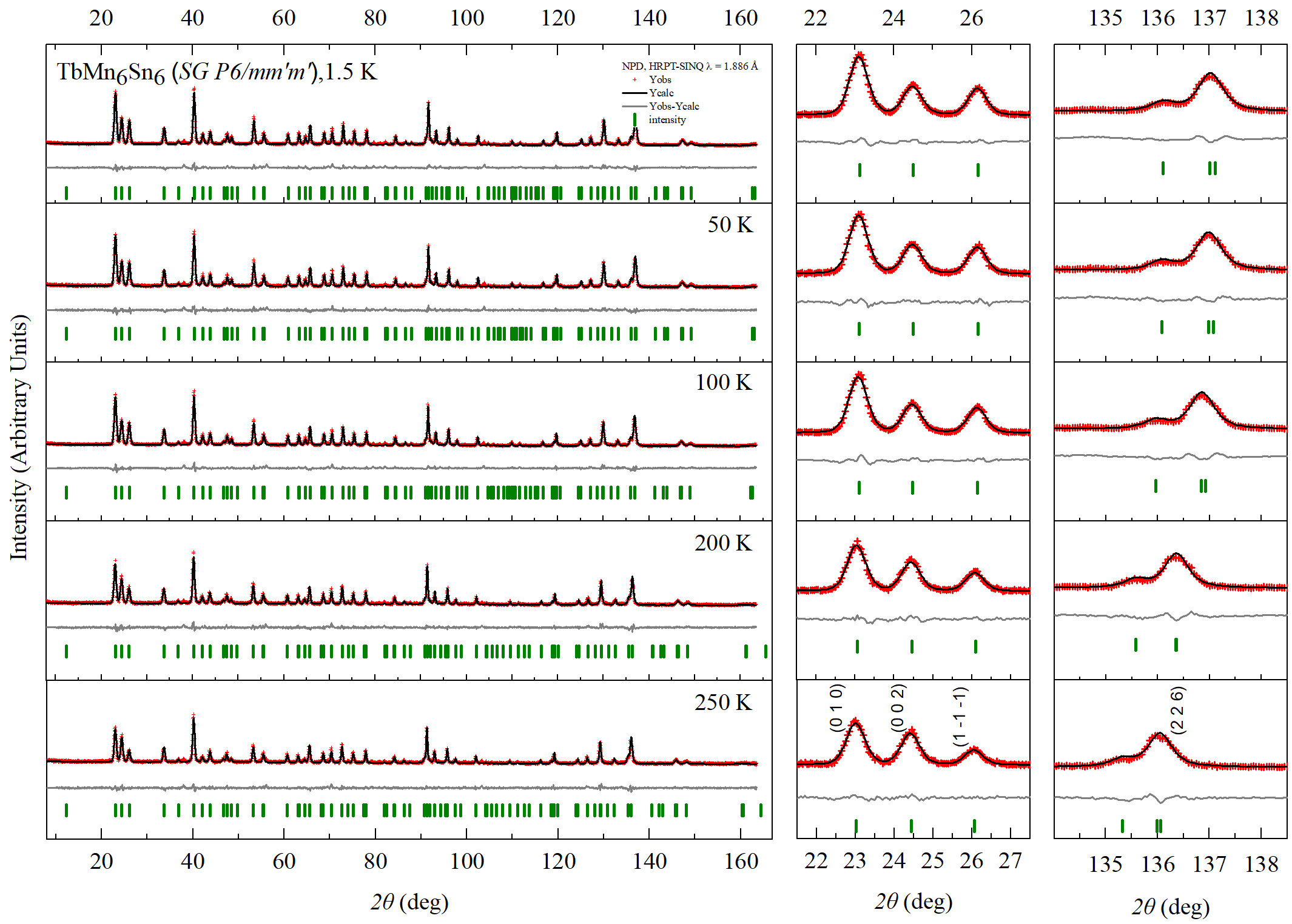}
\vspace{-0.3cm}
\caption{ (Color online) \textbf{NPD temperature evolution of TbMn$_{6}$Sn$_{6}$.} 
High-statistic neutron diffraction acquisitions were made in the temperature range 1.5-250~K. No new peaks are observed over the entire measured temperature range; however, peak intensities smoothly vary with temperature. This indicates that the sample does not undergo a structural phase transition over the entire measured temperature range, and that only the magnitude of the magnetic moments changes with temperature. Several peaks have been highlighted in the insets; we follow the evolution of the (0 1 0), (0 0 2), and (1 -1 -1) peaks to illustrate the change in magnetic moment size and absence of peak splitting. Similarly, we followed the (2 2 6) peak to show the shift in lattice parameters with temperature. Refined values of the lattice parameters and magnetic moments can be found in the main paper, in Figure 2.}
\label{fig2}
\end{figure*}
%%%%%%%%%%%%%%%%%%%%%%%%%%%%%%%%%%%%%%%%%%%%%%%%%%%%%%%%%%%%%%%

%%%%%%%%%%%%%%%%%%%%%%%%%%%%%%%%%%%%%%%%%%%%%%%%%%%%%%%%%%%%%%
\begin{figure*}[t!]
\centering
\includegraphics[width=0.5\linewidth]{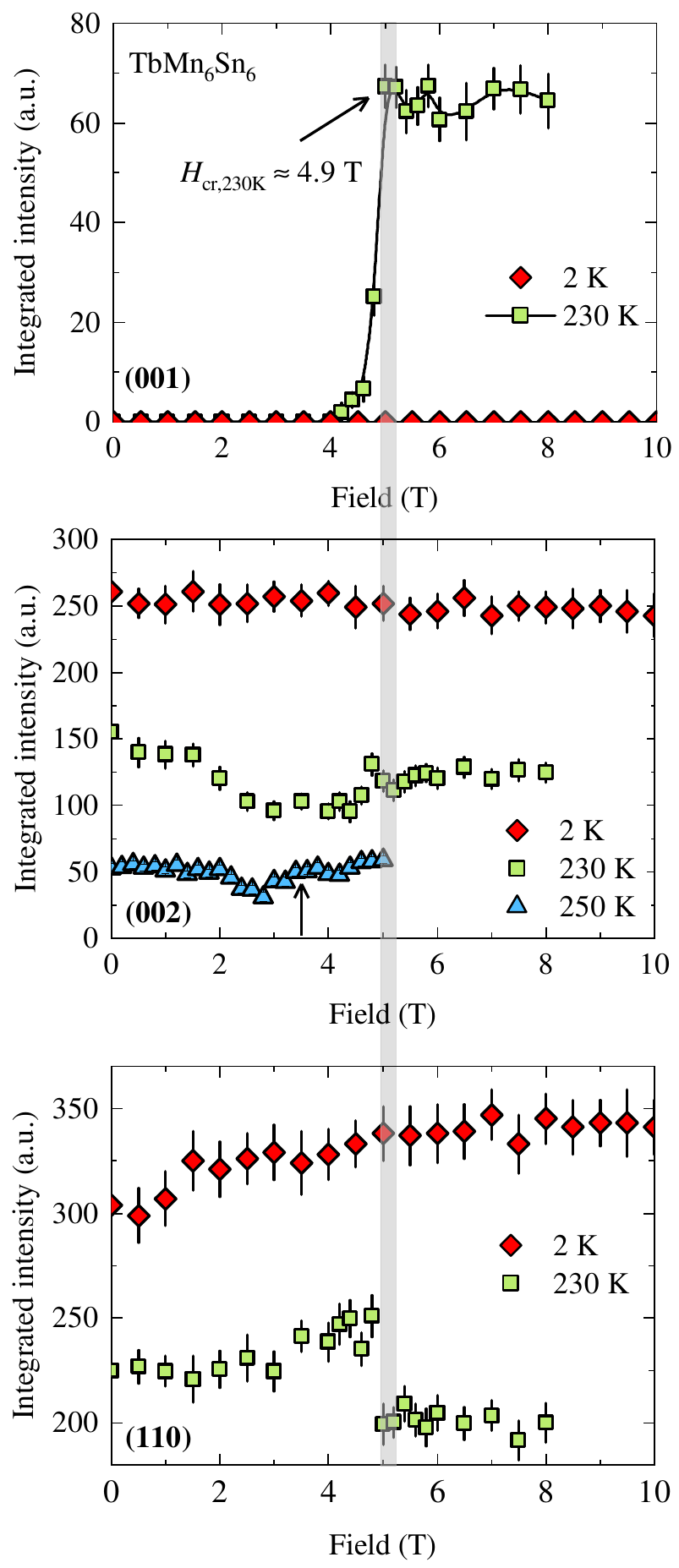}
\vspace{-0.0cm}
\caption{ (Color online) \textbf{Field dependence of single crystal neutron diffraction peak intensities for TbMn$_{6}$Sn$_{6}$.} 
With the single crystal neutron diffractometer ZEBRA (SINQ/PSI), we followed several peak intensities with temperature and applied field. At 230 K, we were able to induce the spin reorientation transition, signaled by the abrupt emergence of the (001) peak at a critical applied field of 4.9 T applied in-plane (top panel). At the base temperature (2 K), however, such a spin reorientation transition was not induced, even with the maximum applied magnetic field of 10 T. This transition was also noticed on the (110) peak (lower panel); however, in the middle panel, no transition was noticed. Coherent with NPD, the (002) peak is a purely nuclear peak, with no magnetic contribution, which is why it is not affected by a change in the magnetic structure induced by applied magnetic field.
}
\label{fig2}
\end{figure*}
%%%%%%%%%%%%%%%%%%%%%%%%%%%%%%%%%%%%%%%%%%%%%%%%%%%%%%%%%%%%%%%

%%%%%%%%%%%%%%%%%%%%%%%%%%%%%%%%%%%%%%%%%%%%%%%%%%%%%%%%%%%%%%
\begin{figure*}[t!]
\centering
\includegraphics[width=1.0 \linewidth]{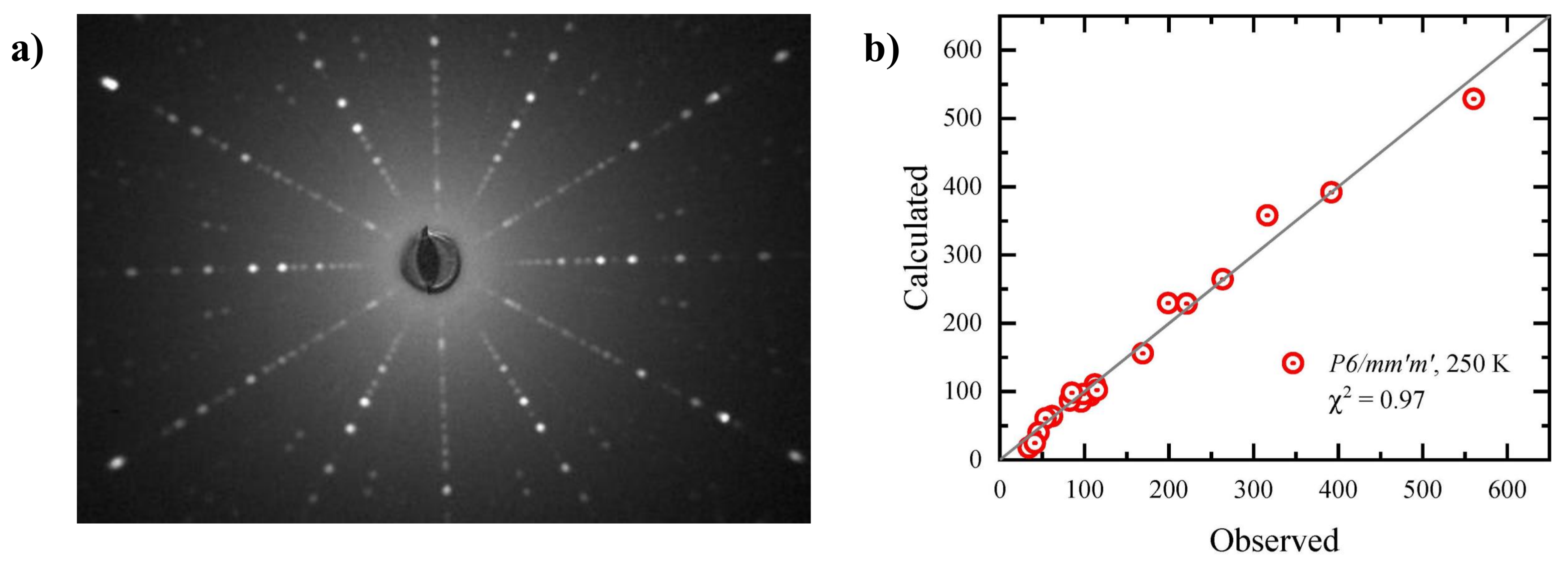}
\vspace{-0.0cm}
\caption{ (Color online) \textbf{Single crystal X-Ray and Neutron Diffraction for TbMn$_{6}$Sn$_{6}$.} 
In figure a) we see the diffraction pattern obtained from Laue X-ray diffraction. This technique was used to select a single crystal sample for neutron diffraction and check for the presence of twins. The platelike crystal was oriented with the thin direction along the X-ray beam; this is the crystallographic $c$-axis. In Figure b) we see the fit from our refinement in $P6/mm'm'$ of the single crystal neutron diffraction data taken at 250 K with $\chi^{2}$ = 0.97. The grey line denotes perfect agreement between observed and predicted intensities.
}
\label{fig2}
\end{figure*}
%%%%%%%%%%%%%%%%%%%%%%%%%%%%%%%%%%%%%%%%%%%%%%%%%%%%%%%%%%%%%%%

\bibliographystyle{prsty}

\end{document}